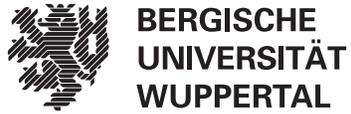

# Accelerating Lattice QCD Simulations using GPUs

Tilmann Matthaei

School of Mathematics and Natural Sciences
University of Wuppertal

August 2023

# Preamble

Solving discretized versions of the Dirac equation represents a large share of execution time in lattice Quantum Chromodynamics (QCD) simulations. Many high-performance computing (HPC) clusters use graphics processing units (GPUs) to offer more computational resources. Our solver program, DDαAMG, previously was unable to fully take advantage of GPUs to accelerate its computations. Making use of GPUs for DDαAMG is an ongoing development, and we will present some current progress herein.

Through a detailed description of our development, this thesis should offer valuable insights into using GPUs to accelerate a memory-bound CPU implementation. Chapter 1 will introduce the basic concepts of lattice QCD, flexible generalized minimal residual (FGMRES) and CUDA. We will also introduce our DDαAMG solver program. Chapter 2 will describe the algorithm currently used to apply the Wilson-Dirac operator. It is that algorithm that we will port to the GPU. Thereafter, Chapter 3 will deal with the performance considerations, which most formed our work on the GPU implementation. The GPU port itself is described in Chapter 4. Before we report our results from the Jülich Wizard for European Leadership Science (JUWELS) HPC cluster in Chapter 6, Chapter 5 goes in detail on competitive configurations for our measurements. We conclude with Chapter 7 by summarizing and discussing our findings. In the same chapter, we will also suggest future research opportunities.

We developed a storage scheme for multiple tuples, which allows much more efficient memory access on GPUs, given that the element at the same index is read from multiple tuples simultaneously. Still, our implementation of a discrete Dirac operator is memory-bound, and we only achieved improvements for large linear systems on few nodes at the JUWELS cluster. These improvements do not currently overcome additional introduced overheads. However, the results for the application of the Wilson-Dirac operator show a speedup of around 3 for large lattices. If the additional overheads can be eliminated in the future, GPUs could reduce the DDαAMG execution time significantly for large lattices.

We also found that a previous publication on the GPU acceleration of DDαAMG, underrepresented the achieved speedup, because small lattices were used. This further highlights that GPUs often require large-scale problems to solve in order to be faster than CPUs.

# List of Acronyms

**AMG** algebraic multigrid

**BiCGSTAB** stabilized biconjugate gradient

**CGN** conjugate gradient on the normal equations

**CPU** central processing unit

**DD** domain decomposition

**FGMRES** flexible generalized minimal residual

**GMRES** generalized minimal residual

**GPU** graphics processing unit

**HBM** High Bandwidth Memory

**HPC** high-performance computing

**JUWELS** Jülich Wizard for European Leadership Science

**MPI** message passing interface

**NUMA** Non-Uniform Memory Access

**PCIe** Peripheral Component Interconnect Express

**PDE** partial differential equation

**PTX** parallel thread execution

**QCD** Quantum Chromodynamics

**SAP** Schwarz alternating procedure

**SIMT** Single Instruction Multiple Thread

**SM** Streaming Multiprocessor

**SSE** Streaming SIMD Extensions

# Mathematical Notation

$\begin{bmatrix} a \\ b \\ c \end{bmatrix} = (a, b, c)$  Column Vector

$A^H$  Hermitian Transpose of $A$.
$A \otimes B$  Kronecker Product.
$\|x\|$  (Unspecified) Norm of vector $x$.
$\|x\|_2$  Euclidean Norm / 2-Norm of vector $x$.
$|A|_{card}$  Cardinality of a Set $A$
$s_*$  An indexed symbol $s$ where in the specific case the index is of no importance.
$\mathbb{N}$  Natural Numbers, including 0
$\mathbb{N}^+$  Natural Numbers, excluding 0

# Contents

# Chapter 1

# Introduction

This chapter will introduce basic concepts that will be required to understand the upcoming chapters. We will give a basic introduction to lattice Quantum Chromodynamics (QCD). We will also introduce flexible generalized minimal residual (FGMRES), a commonly used iterative solver for linear systems. DDαAMG is our solver program for discretized versions of the Dirac equation and will also be presented. Finally, the CUDA framework will be introduced and basic concepts of CUDA programming will be explained. None of these sections can give a complete description of their respective subjects, and the reader is encouraged to investigate them further using cited literature.

## 1.1 Lattice QCD and the Wilson-Dirac Operator

"Lattice QCD is a framework in which the theory of strong interactions can be studied from first principles." [Bla+22]

QCD is the physical theory, that describes the interactions of quarks as the constituents of matter. This strong interaction is one of four fundamental forces in physics, the three other ones being the gravitational force, the weak force and the electromagnetic force. At this point, a full introduction to the theory on the strong interactions between quarks and gluons is out of scope. That is mostly due to the fact that this theory is not relevant to the optimizations that are the topic of this thesis. This section will rely heavily on the information and conventions from [Fro+14].

Some predictions in QCD require the theory of QCD to be discretized and simulated numerically. An important and computationally expensive task in this process is the solution of the discretized Dirac equation for a given right-hand side. To properly introduce the discretization scheme of the Dirac equation, we will now give a basic definition of it in the continuum.

A spinor in QCD is a twelve component complex vector $s \in \mathbb{C}^{12}$, which can be indexed by color and spin. Color indices $\mathcal{C}$ and spin indices $\mathcal{S}$ are

$$\mathcal{C} = \{1, 2, 3\}, \mathcal{S} = \{0, 1, 2, 3\}. \tag{1.1}$$

The (typically suppressed) indexing scheme is given by

$$s = s_{c,\sigma} := (s_{1,0}, s_{2,0}, s_{3,0}, s_{1,1}, \ldots, s_{3,3}), \tag{1.2}$$

and

$$s_\sigma := (s_{1,\sigma}, s_{2,\sigma}, s_{3,\sigma}). \tag{1.3}$$

In the continuum case, quark fields $\psi$ and $\eta$ are mappings between (continuous) points in spacetime $\mathbb{R}^4$ and spinors. We now introduce the continuum Dirac equation:

$$(\mathcal{D} + m)\psi = \eta, \tag{1.4}$$

where $m$ is a scalar mass parameter not depending on $x$.

We define the quark fields $\psi$ and $\eta$ as

$$\begin{aligned} \psi &: \mathbb{R}^4 \to \mathbb{C}^{12}, \quad x \mapsto \psi(x) \text{ and} \\ \eta &: \mathbb{R}^4 \to \mathbb{C}^{12}, \quad x \mapsto \eta(x). \end{aligned} \tag{1.5}$$

The Dirac operator $\mathcal{D}$ can be written as

$$\mathcal{D} = \sum_{\mu=0}^{3} \gamma_\mu \otimes (\partial_\mu + A_\mu), \tag{1.6}$$

where $\partial_\mu = \partial/\partial x_\mu$ and $A_\mu$ is the gauge field. For details on the properties of the continuum Dirac operator we refer the reader to [Fro+14].

A lattice, as used throughout this thesis, is a collection of per-direction equidistant points in spacetime. These points are also referred to as lattice sites. The underlying concept or "type" of these points will be largely irrelevant. It shall be mentioned that the previous work by Rottmann [Rot16] considered $\mathcal{L} \subset \mathbb{R}^4$ to be a finite set of vectors from $\mathbb{R}^4$ representing a discretization of spacetime. That is particularly convenient when working with the integrals that are required when deducing lattice QCD theory from continuum QCD. Implementation-wise, it suffices to consider $\mathcal{L} \subset \mathbb{N}$ to be a set of one-dimensional indices referring to these points. This brings the formulation of related algorithms closest to their actual implementation. Within the scope of this introduction we will consider a four-dimensional indexing scheme.

We define the direction of time $T$ and 3 directions of space $Z, Y, X$ to be

$$T = \begin{bmatrix} 1 \\ 0 \\ 0 \\ 0 \end{bmatrix}, Z = \begin{bmatrix} 0 \\ 1 \\ 0 \\ 0 \end{bmatrix}, Y = \begin{bmatrix} 0 \\ 0 \\ 1 \\ 0 \end{bmatrix}, X = \begin{bmatrix} 0 \\ 0 \\ 0 \\ 1 \end{bmatrix}. \tag{1.7}$$

The canonical basis of $\mathbb{N}^4$ thus represents directions of spacetime.

$$\mathbb{D} := \{T, Z, Y, X\} \tag{1.8}$$

The size of the lattice in each direction is given by $n_T, n_Z, n_Y, n_X \in \mathbb{N}^+$.

The lattice $\mathcal{L}$ is defined as

$$\mathcal{L} := \begin{matrix} \{1,\ldots,n_T\} \times \{1,\ldots,n_Z\} \times \\ \{1,\ldots,n_Y\} \times \{1,\ldots,n_X\} \end{matrix} \subset \mathbb{N}^{+4}. \tag{1.9}$$

The total size of the lattice is $n_{\mathcal{L}} := |\mathcal{L}|_{card} = n_T n_Z n_Y n_X$.

A lattice site is denoted as $x \in \mathcal{L}$. We refer to $x + \mu$ as the neighbor of $x$ in positive $\mu$ direction, and we refer to $x - \mu$ as the neighbor of $x$ in negative $\mu$ direction. As we study the Dirac equation with (anti-)periodic boundary conditions, we consider evaluations of $U_\mu, \psi$ or $\eta$ where $x \pm \mu \notin \mathcal{L}$ to be given by an appropriate lattice site (e.g. $\psi((n_T, x_2, x_3, x_4) + T) \sim \psi((1, x_2, x_3, x_4)))$. We consider lattice sites, which we can use as a substitute for $x \pm \mu$ in such cases, neighbors of $x$ in positive/negative $\mu$ direction as well. We designate the distance between neighboring points in the physical system as $a$ (disregarding units).

To illustrate some concepts throughout this thesis, we use figures that simplify the structure to a two-dimensional lattice. We use $\mathcal{L} \subset \mathbb{N}^{+2}$ accordingly. Figure 1.1 helps us to visualize the structure of the lattice in a two-dimensional setting.

We will now construct the Wilson discretization of the continuum Dirac operator from Equation 1.6. In the discrete case, quark fields $\psi, \eta : \mathcal{L} \to \mathbb{C}^{12}$ are maps between lattice sites and spinors.

A discrete gauge configuration $U = \{U_\mu(x), x \in \mathcal{L}, \mu \in \mathbb{D}\}$ shall be given and can be derived from the physical gauge field. For details on the derivation of the gauge configuration we refer the reader to [MM94, pp. 97 ff.]. A matrix $U_\mu(x)$ is also called a gauge link. The matrices $U_\mu(x)$ satisfy $U_\mu(x) \in SU(3) \Rightarrow U_\mu^{-1}(x) = U_\mu^H(x)$, where $SU(3)$ is the special unitary group of degree 3 (i.e. the Lie group of $n \times n$ unitary matrices with determinant 1).

We thus construct forward covariant finite differences

$$(\Delta^\mu \psi_\sigma)(x) := \frac{U_\mu(x)\psi_\sigma(x+\mu) - \psi_\sigma(x)}{a},$$

backward covariant finite differences

$$(\Delta_\mu \psi_\sigma)(x) := \frac{\psi_\sigma(x) - U_\mu^H(x-\mu)\psi_\sigma(x-\mu)}{a}$$

and a centralized covariant finite difference discretization of the (continuous) Dirac operator $\mathcal{D}$

$$D_N = \sum_{\mu \in \mathbb{D}} \gamma_\mu \otimes \frac{1}{2}(\Delta_\mu + \Delta^\mu).$$

The stabilization term $a\Delta_\mu \Delta^\mu$ is a centralized second order covariant finite difference. Given a mass parameter $m_0$, the Wilson discretization of the Dirac operator (also Wilson-Dirac operator) is then defined by

$$D_W = \frac{m_0}{a}I + \frac{1}{2}\sum_{\mu \in \mathbb{D}}(\gamma_\mu \otimes (\Delta_\mu + \Delta^\mu) - aI_4 \otimes \Delta_\mu \Delta^\mu). \tag{1.10}$$

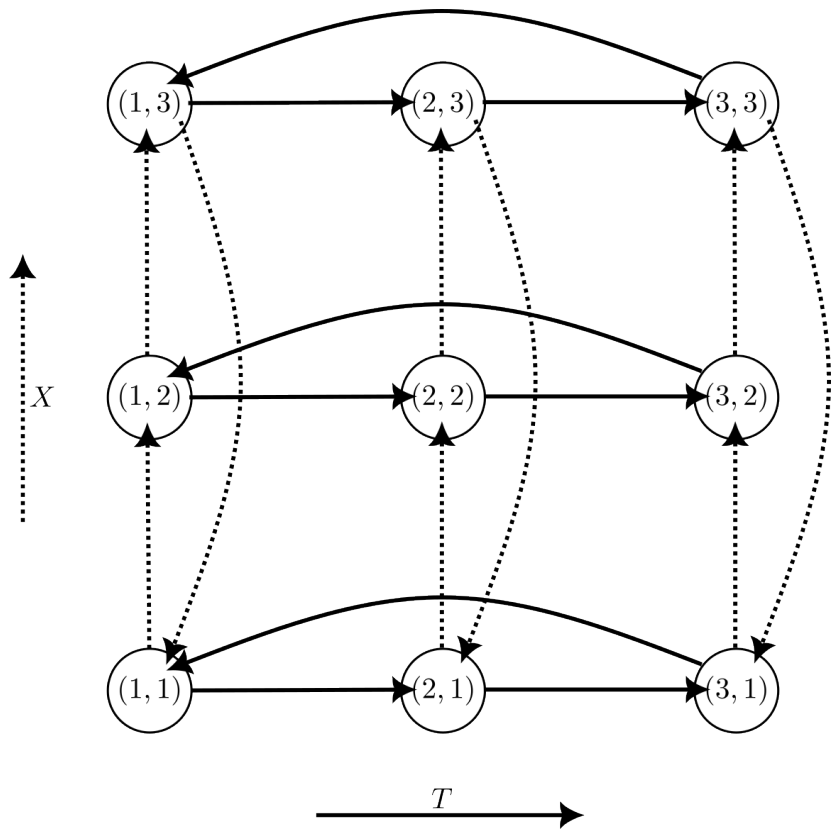

Figure 1.1: 2D Lattice

To reduce the order of the discretization error as a function of $a$, the Sheikholeslami-Wohlert or "clover" term, depending on a parameter $c_{sw}$, is added to the lattice Wilson-Dirac operator. This term was introduced in [SW85]. It removes $O(a)$ discretization errors, while maintaining the advantageous nearest-neighbor coupling:

$$D = D_W - \frac{c_{sw}}{32a} \sum_{\mu,\nu \in \mathbb{D}^2} (\gamma_\mu \gamma_\nu) \otimes (Q_{\mu\nu} - Q_{\nu\mu}) \tag{1.11}$$

where $(Q_{\mu\nu}\psi_\sigma)(x) = Q_{\mu\nu}(x)\psi_\sigma(x)$ with

$$\begin{aligned}
Q_{\mu\nu}(x) &= U_\mu(x)U_\nu(x+\mu)U_\mu(x+\nu)^H U_\nu(x)^H \\
&\quad + U_\nu(x)U_\mu(x-\mu+\nu)^H U_\nu(x-\mu)^H U_\mu(x-\mu) + \\
&\quad + U_\mu(x-\mu)^H U_\nu(x-\mu-\nu)^H U_\mu(x-\mu-\nu)U_\nu(x-\nu) \\
&\quad + U_\nu(x-\nu)^H U_\mu(x-\nu)U_\nu(x-\nu+\mu)U_\mu(x)^H.
\end{aligned} \tag{1.12}$$

The formulation of D can be rearranged, such that the terms including $\psi(x)$, $\psi(x+\mu)$ and $\psi(x-\mu)$ are separated (see [Fro+14]).

$$\begin{aligned}
(D\psi)(x) &= \frac{1}{a}\left((m_0 + 4)I_{12} - \frac{c_{sw}}{32}\sum_{\mu,\nu \in \mathbb{D}^2}(\gamma_\mu\gamma_\nu) \otimes (Q_{\mu\nu}(x) - Q_{\nu\mu}(x))\right)\psi(x) \\
&\quad - \frac{1}{2a}\sum_{\mu \in \mathbb{D}}((I_4 - \gamma_\mu) \otimes U_\mu(x))\psi(x+\mu) \\
&\quad - \frac{1}{2a}\sum_{\mu \in \mathbb{D}}((I_4 + \gamma_\mu) \otimes U_\mu^H(x-\mu))\psi(x-\mu).
\end{aligned} \tag{1.13}$$

## 1.2 GMRES and FGMRES

Generalized minimal residual (GMRES) is an iterative solver for linear systems $Ax = b$. It is a projection method based on Krylov subspaces. For the $m$-th Krylov subspace, GMRES minimizes the residual norm over all vectors in $x_0 + \mathcal{K}_m$. The iteration count of these methods depends on the conditioning of the linear operator $A$. Therefore, often preconditioning is applied to reduce the iteration count. Flexible iterative methods, such as FGMRES, allow to apply a different preconditioning each iteration. Popular choices for preconditioning are other iterative methods, which we will discuss in Chapter 1.3. Such an approach inherently requires flexible preconditioning.

The basic algorithm for FGMRES may be found in [Saa93]. The algorithm will be repeated here as Algorithm 2 in slightly altered form (but functionally equivalent), in order to highlight where the GPU acceleration will take place. For comparison, we also provide the GMRES algorithm with right preconditioning as Algorithm 1.

The termination criterion in DDαAMG is that either the residual is below a configurable tolerance $\|b - Ax\| < tol\, \|b - Ax_0\|$ or that a configurable number of iterations

**Algorithm 1:** GMRES with right Preconditioning

**Input:** Linear operator $A$, inverse of the preconditioner $M^{-1}$, right-hand side $b$, dimension $m$ of the Krylov subspaces.
**Output:** Approximate solution $x$ to the linear system $Ax = b$.

1 Make an initial guess $x \leftarrow x_0$;
2 $H_m \in \mathbb{C}^{(m+1) \times m} \leftarrow 0^{(m+1) \times m}$;
  /* Outer GMRES (restart) loop                          */
3 **while** *termination criterion not satisfied* **do**
    /* Arnoldi Process                                   */
4     $r \leftarrow b - Ax$;
5     $\beta \leftarrow \|r\|_2$;
6     $v_1 \leftarrow \frac{1}{\beta} r$;
7     **for** $j \leftarrow 1$ **to** $m$ **do**
8       $z_j \leftarrow M^{-1} v_j$;
9       $w \leftarrow A z_j$;
10       **for** $i \leftarrow 1$ **to** $j$ **do**
11         $h_{i,j} \leftarrow \langle w, v_i \rangle$;
12         $w \leftarrow w - h_{i,j} v_i$;
13       **end**
14       $h_{j+1,j} \leftarrow \|w\|_2$;
15       $v_{j+1} \leftarrow \frac{1}{h_{j+1,j}} w$;
16     **end**
17     $V_m \leftarrow [v_1, \ldots, v_m]$;
    /* Form the approximate solution                     */
18     $y_m \leftarrow \mathrm{argmin}_y \|\beta e_1 - H_m y\|_2$;
19     $x_m \leftarrow x + M^{-1} V_m y_m$;
20     $x \leftarrow x_m$
21 **end**
22 **return** $x$

have been performed. Additionally, DDαAMG terminates early if $|h_{j+1,j}| < \frac{tol}{10}$, which indicates that the method is near breakdown and $x_m$ is close to an exact solution (see [Saa03, p. 179]).

The FGMRES method introduces some small changes to allow the preconditioner to vary between iterations. That is, the vectors $z_j$ now need to be saved in order to form the approximate solution. The resulting FGMRES algorithm is given by Algorithm 2. It is also an option to save the vectors $z_j$ with a constant preconditioner as a variant of Algorithm 1. This is a time-memory trade-off.

---

**Algorithm 2:** FGMRES

---

**Input:** Linear operator $A$, inverse of the preconditioner $M_j^{-1}$, right-hand side $b$, dimension $m$ of the Krylov subspaces.

**Output:** Approximate solution $x$ to the linear system $Ax = b$.

1 Make an initial guess $x \leftarrow x_0$.;
2 $H_m \in \mathbb{C}^{(m+1) \times m} \leftarrow 0^{(m+1) \times m}$;
   /* Outer FGMRES (restart) loop                                       */
3 **while** *termination criterion not satisfied* **do**
     /* Arnoldi Process                                                    */
4     $r \leftarrow b - Ax$;
5     $\beta \leftarrow \|r\|_2$;
6     $v_1 \leftarrow \frac{1}{\beta} r$;
7     **for** $j \leftarrow 1$ **to** $m$ **do**
8         $z_j \leftarrow M_j^{-1} v_j$;
9         $w \leftarrow A z_j$;
10        **for** $i \leftarrow 1$ **to** $j$ **do**
11            $h_{i,j} \leftarrow \langle w, v_i \rangle$;
12            $w \leftarrow w - h_{i,j} v_i$;
13        **end**
14        $h_{j+1,j} \leftarrow \|w\|_2$;
15        $v_{j+1} \leftarrow \frac{1}{h_{j+1,j}} w$;
16     **end**
17     $Z_m \leftarrow [z_1, \ldots, z_m]$;
     /* Form the approximate solution                          */
18     $y_m \leftarrow \operatorname{argmin}_y \|\beta e_1 - H_m y\|_2$;
19     $x_m \leftarrow x + Z_m y_m$;
20     $x \leftarrow x_m$
21 **end**
22 **return** $x$

---

The linear operator $A$ that is being studied in the DDαAMG project is the Wilson-Dirac operator. In the FGMRES Algorithm 2, this operator is applied in line 4 once per outer iteration and then $m$ times per outer iteration in line 9. The topic of this thesis is the GPU acceleration of the application of this Wilson-Dirac operator as required for the FGMRES algorithm.

FGMRES (Algorithm 2) is equivalent to GMRES with right preconditioning (Algorithm 1) in the case that the preconditioner does not change between iterations (e.g. $M_j = M$) as discussed in [Saa93]. That in turn is equivalent to standard GMRES provided the preconditioner is the identity operator $I$. Both no preconditioning (standard GMRES) and non-stationary preconditioning are possible with DDαAMG depending on its configuration. Application of the operator is required whichever of those algorithms is used. Only the parameters $x$ and $z_j$ are different and the computational effort of applying the Wilson-Dirac operator (once) is the same over all variants. This has two important implications for efforts regarding the acceleration of the application of the operator. Firstly, the performance impact can be measured independently of whether FGMRES or GMRES (with or without preconditioner) is used. Secondly, the improvements take effect throughout all solver configurations of DDαAMG.

## 1.3 The DDαAMG project

DDαAMG is a solver program for large sparse linear systems of equations $Dz = b$ stemming from the Wilson discretization $D$ of the Dirac operator. It is written in C with some functionalities already having alternative implementations in CUDA C++. Message passing interface (MPI) is used as a communication framework and the solver program is thus able to run on distributed systems including high-performance computing (HPC) clusters.

DDαAMG has been first introduced in the conference proceedings [Rot+12]. The article describes the basic approach of DDαAMG, which combines a domain decomposition (DD) and algebraic multigrid (AMG) method. In 2014, it became the topic of a subsequent journal article [Fro+14]. The doctoral thesis [Rot16] contains the most complete description of the algorithm used to apply the Wilson-Dirac operator on the finest level (pp. 29-32). It is that specific algorithm whose port to the graphics processing unit (GPU) is the topic of this thesis. The algorithm is discussed in detail in Chapter 2.

DDαAMG implements multiple solving methods for cross-comparison, namely conjugate gradient on the normal equations (CGN), GMRES and FGMRES. Within FGMRES it allows the use of an AMG method as a preconditioner. For the AMG method, the additive, red-black or 16 color multiplicative Schwarz methods, GMRES or stabilized biconjugate gradient (BiCGSTAB) method can be used as a smoother. This multigrid method acts as a (right) preconditioner in the sense of FGMRES. All mentioned algorithms are described in [Saa03].

The most notable achievements of the DDαAMG project come from its AMG preconditioner. The AMG method drastically lowers the amount of FGMRES iterations required to solve the linear system. While DDαAMG's preconditioning generally targets ill-conditioned systems, the AMG method can even solve a fairly well-conditioned system with significantly fewer (finest-level Wilson-Dirac) operator applications compared to non-preconditioned GMRES. Table 6.3 shows an example where the AMG-preconditioned FGMRES requires 8 instead of 124 operator applications and FGMRES did not need to be restarted.

Before the introduction of the AMG method, we will discuss the usefulness of an (approximate) solver as a preconditioner in FGMRES. Assuming the inverse $A^{-1}$ of the linear operator $A$ exists and is known. $\kappa(A^{-1}A) = \kappa(I) = 1$ shows the optimality of the inverse of a matrix as a preconditioner. Analytically, we find that left preconditioning $Ax = b$ with $A^{-1}$ results in $x = A^{-1}b$, i.e. we obtain $x$ directly. When right preconditioning $Ax = b$ with $A^{-1}$ we obtain the system in $AA^{-1}(Ax) = Iy = b$ and $Ax = y$ can be solved through our knowledge of $A^{-1}$.

We now verify that $A^{-1}$ makes a good preconditioner $M_j^{-1}$ for FGMRES in Algorithm 2. The preconditioner is first applied in line 8. As $v_1 = \frac{1}{\beta}r$ in the first iteration we obtain $z_1 \leftarrow \frac{1}{\beta}A^{-1}r = \frac{1}{\beta}(A^{-1}b - x_0)$ and $w \leftarrow \frac{1}{\beta}r$. The subsequent loop only has one iteration and calculates $h_{1,1} = \langle w, v_1 \rangle = \left\langle \frac{1}{\beta}r, \frac{1}{\beta}r \right\rangle = 1$ ($\frac{1}{\beta}r$ is a unit vector) as well as $w \leftarrow w - h_{1,1}v_1 = v_1 - 1v_1 = 0$. Finally, $h_{j+1,j} \leftarrow \|w\|_2 = 0$ and the method breaks down in line 15 due to a division by zero forcing an early stop.

To build the correction, $Z_m \leftarrow [z_1] = [\frac{1}{\beta}(A^{-1}b - x_0)]$, and then

$$y_m \leftarrow \underset{y}{\operatorname{argmin}} \left\| \beta e_1 - \begin{bmatrix} 1 & 0 & \cdots \\ 0 & 0 & \\ \cdots & & \ddots \end{bmatrix} y \right\|_2 = \beta e_1.$$

So, the approximate solution $x_m \leftarrow x_0 + (A^{-1}b - x_0) = A^{-1}b$ is in fact the exact solution, which was obtained through a single inner iteration. The preconditioner $A^{-1}$ is an optimal preconditioner to FGMRES, in the sense that FGMRES converges in a single (inner Arnoldi) iteration.

In practice, the inverse $A^{-1}$ is not known and computing it exactly is as expensive as solving the original system. However, under certain restrictions an approximation of the inverse can also be a good preconditioner. This gives rise to the class of Approximate Inverse Preconditioners [Saa03; CS94]. DDαAMG does not construct an inverse as a matrix, but rather uses another iterative method as a preconditioner. The AMG method should conceptually provide an approximate solution to the linear system $Az_j = v_j$, that reduces residual components not easily reduced by FGMRES.

We refer to the space spanned by the eigenvectors belonging to small (absolute) eigenvalues of $D$ as *near kernel*. The Schwarz alternating procedure (SAP) and many other iterative methods are not able to reduce error components belonging to the near kernel sufficiently. Details on this issue are discussed in [Fro+14, p. 9] and [Saa03, pp. 423 ff.]. The AMG method developed for the DDαAMG solver makes a projection of the Wilson-Dirac operator $D$ into a smaller ("coarser") grid. The so-created linear operator $D_c$ acts on the near kernel (of $D$) similarly to $D$ and also preserves the connection structure and sparsity of $D$. That is, $D_c$ also only couples nearest neighbors. The linear restriction map $R: \mathbb{C}^n \to \mathbb{C}^{n_c}$ and linear prolongation map $P: \mathbb{C}^{n_c} \to \mathbb{C}^n$ allow the transfer of information from and to the finer grid/lattice. The subspace correction, $z \leftarrow z + PD_c^{-1}Rr$, is expected to effectively reduce components of the iterate from the smoother belonging to the near kernel. Together with the smoother we obtain an approximate solution to $Dz_j = v_j$ and thus a good preconditioning for FGMRES.

For the construction of $P$ and $R$, approximations of right eigenvectors belonging to small eigenvalues of $D$ need to be obtained. DDαAMG therefore has an expensive setup phase to obtain those approximations. For systems that could be solved by standard Krylov subspace methods that setup cost often does not pay off (see Table 6.4). If one would reuse the obtained $P$, $R$ and $D_c$ to solve multiple right-hand sides the setup cost might pay off eventually. This feature is however not yet supported by the DDαAMG solver. A good choice of multigrid parameters can also balance the cost of the preconditioner against the FGMRES iteration reduction.

## 1.4 CUDA C++

This section is heavily based on the *CUDA C++ Programming Guide* published by NVIDIA [NVI23]. As this is a manufacturer-supplied manual, it does not conform to scientific standards and, quite on the contrary, the manufacturer could be enticed to inflate statements about the capability and maturity of its product. However, with CUDA being proprietary software, that manual is also the most up-to-date and complete description of the CUDA model and interface and has thus been included in the bibliography.

In 2006 NVIDIA introduced CUDA [NVI23, p. 5], a computation platform that allows use of CUDA-capable GPUs for general-purpose computation tasks. CUDA C++ is a "set of extensions to the C++ language and a runtime library" [NVI23, p. 21]. Other compilers do no understand this dialect of C++ and source code must be compiled by the NVIDIA-supplied `nvcc` compiler. CUDA C++ source files can contain a mix of host code (code that is being executed by the central processing unit (CPU)) and so-called kernels. Kernels are functions that can be executed by the GPU in parallel. Host code calls kernels with an execution configuration that specifies the amount of simultaneous threads that execute that kernel. Threads are organized in blocks (of threads) and a grid (of blocks). Each thread receives a (possibly multidimensional) block and grid index.

Threads within the same block are processed in groups of 32, called warps. All threads within a warp execute one common instruction at a time. Their operands are allowed to be different without loss of performance, but conditional branches are executed sequentially. Threads that do not execute that branch become idle, which is known as warp divergence (which is almost always undesirable). NVIDIA calls this architecture Single Instruction Multiple Thread (SIMT). This architecture gives raise to multiple hardware-level optimization opportunities (e.g. coalescing of memory requests or optimized scheduling). Multiple warps can be "active" on each Streaming Multiprocessor (SM), the processing elements of warps. That is, warps can be swapped in and out of execution to hide latency, similar to simultaneous multithreading technologies on CPUs.

The GPU has its own completely separate memory and L1 and L2 cache. L2 cache is shared between SMs, while L1 cache is exclusive to each SM. NVIDIA refers to the GPU main memory as *global memory*. Through the GPUDirect RDMA technology, the contents of GPU memory can be directly transferred to another GPU or another PCIe device (e.g. an InfiniBand adapter). This transfer can be performed asynchronously without blocking the CPU. The L1 cache can be partially manually managed through *shared memory*. That is, a portion of L1 cache can be allocated for use within a thread block. Data can be transferred between CPU memory and GPU memory through the Peripheral Component Interconnect Express (PCIe) bus.

Ramírez-Hidalgo, another developer of DDαAMG, already ported its finest-level SAP smoother to CUDA C++. In [Ram22, p. 93], he reports a speedup of more than 30 for the computational work associated with the smoother and a total smoother speedup of more than 20. However, total execution time in the booster module for his numerical experiments on the JUWELS supercomputer at the Jülich Supercomputing Centre *increased* (from 6.14 seconds to 7.73 seconds) [Ram22, p. 96]. The reason for this is not explicitly discussed, but there are overheads involved in using CUDA C++. Notably, it introduces the requirement to copy data between CPU and GPU over the PCIe bus [McK14, p. 26 ff.]. Ramírez-Hidalgo thus proposes to:

»Port the whole finest level, and not only the smoother at that level. The finest level is very rich in floating point operations, and communications are nicely hidden behind those. By merging the smoother at the finest level with all the other operations there, we can avoid frequent and large data transfers from CPU to GPU and viceversa.«[Ram22, p. 96 f.]

The developers of the GPU-accelerated QUDA library have also adopted a multigrid approach to lattice QCD. Even though they state that "GPUs represent an extremely challenging architecture on which to deploy an efficient MG algorithm" [Cla+16], they report to be ultimately successful in outperforming BiCGSTAB by a factor of (typically) $5 - 8\times$ using GPUs and a multigrid solver.

# Chapter 2

# Finest Level Operator Algorithm

## 2.1 Local Lattices & Processes

As DDαAMG can be run on distributed systems, each process takes care of a subset of the lattice. The set of all lattice sites is referred to as the global lattice and denoted by $\mathcal{L}$ as in the introductory Section 1.1. The subset that is treated by each process is called the local lattice and denoted by $P_i$. Each local lattice is a hyper-rectangular subset of the global lattice. Local lattices do not overlap (i.e. $\forall i, j : P_i \cap P_j = \emptyset$).

The original definition of ghost cells/outer boundaries from [Rot16] is

$$\partial_\mu^\pm P_i := \{x \in \mathcal{L} \setminus P_i : \exists \mu \in \mathbb{D}, x + \mu \in P_i \text{ or } x - \mu \in P_i\}, \quad (2.1)$$

and inner boundaries were defined as

$$d_\mu^\pm P_i := \{x \in P_i : \exists \mu \in \mathbb{D}, x + \mu \in \partial_\mu^\pm P_i \text{ or } x - \mu \in \partial_\mu^\pm P_i\}. \quad (2.2)$$

However, we found that $\partial_\mu^- P_i$ and $d_\mu^+ P_i$ are unused in Algorithm 3 (the application of the discrete Wilson-Dirac operator on the finest level). Due to the computation of $\chi_\mu$ by another process, we do not need to store the values of $\psi$ for lattice sites in $\partial_\mu^- P_i$. As the distinction of positive or negative boundaries thus becomes superfluous, we will drop the $\pm$ index and use

$$\partial_\mu P_i = \partial_\mu^+ P_i \quad (2.3)$$

$$d_\mu P_i = d_\mu^- P_i \quad (2.4)$$

Figure 2.1 shows these boundaries in a simplified two-dimensional way. Clearly, the outer boundaries $\partial_\mu P_i$ are also the inner boundaries of other local lattices. We designate the set of all local lattice sites and all outer boundary lattice sites for process $i$ as

$$\overline{P_i} = P_i \cup \bigcup_{\mu \in \mathbb{D}} \partial_\mu P_i.$$

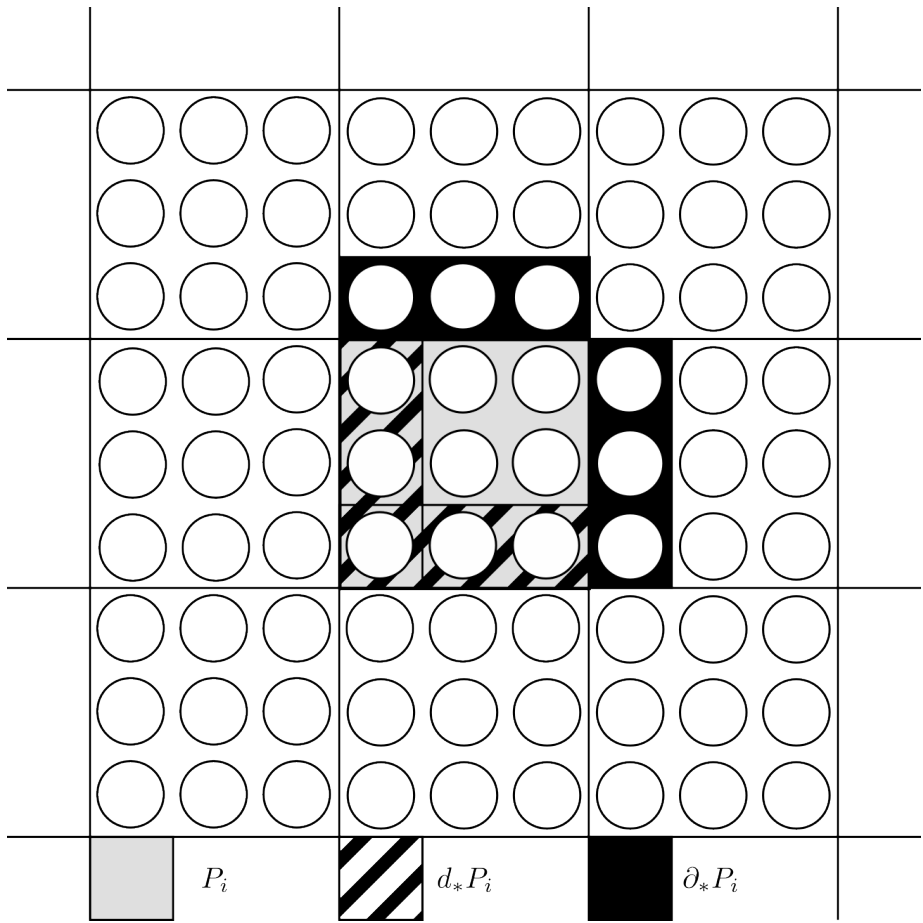

Figure 2.1: Inner and Outer Boundaries of a Local Lattice $P_i$

Accordingly, for a map (e.g. a quark field) $\psi$, we define the inner and outer boundaries of $\psi$ as

$$\partial_{\mu,P_i}\psi := \{\psi(x) : x \in \partial_\mu P_i\} \text{ and } d_{\mu,P_i}\psi := \{\psi(x) : x \in d_\mu P_i\}.$$

As the discretized partial differential equation (PDE) has (anti-)periodic boundary conditions, this creates a "seam" of the outer boundary of one local lattice and the inner boundary of exactly one other local lattice. The two local lattices need not necessarily be distinct. On these "seams" communication with another process that holds a local lattice that is also part of the "seam" will be necessary as will be obvious from the upcoming Algorithm 3. The coupling of nearest neighbors has also been discussed in the introduction of the Wilson-Dirac operator in Section 1.1.

## 2.2 Neighbor Coupling

Equation 1.13 can be separated into the term that is diagonal, i.e. a multiplication with $\psi(x)$, and the terms which are multiplications with $\psi(x+\mu)$ and $\psi(x-\mu)$. We now introduce $D_{sc}$ (**s**elf **c**oupling) and $D_{nc}$ (**n**eighbor **c**oupling):

$$(D_{sc}\psi)(x) = \left((m_0 + 4)I_{12} - \frac{c_{sw}}{32} \sum_{\mu,\nu \in \mathbb{D}^2} (\gamma_\mu \gamma_\nu) \otimes (Q_{\mu\nu}(x) - Q_{\nu\mu}(x))\right)\psi(x) \tag{2.5}$$

$$(D_{nc}\psi)(x) = \frac{1}{2} \sum_{\mu \in \mathbb{D}} ((I_4 - \gamma_\mu) \otimes U_\mu(x))\psi(x + \mu)$$
$$+ \frac{1}{2} \sum_{\mu \in \mathbb{D}} ((I_4 + \gamma_\mu) \otimes U_\mu^H(x - \mu))\psi(x - \mu) \tag{2.6}$$

Now we can rewrite $D$ as a combination of $D_{sc}$ and $D_{nc}$.

$$a(D\psi)(x) = (D_{sc}\psi)(x) - (D_{nc}\psi)(x) \tag{2.7}$$

The matrix for the clover term for each lattice site can be calculated once and does not need to be recalculated for each application of the operator. Calculation of $(D_{sc}\psi)(x)$ is therefore a relatively straightforward matrix-vector-multiplication and matrix addition (which can be fused). This calculation does never depend on values of $\psi$, which are not in the local lattice. Therefore, calculation of the self coupling term $D_{sc}$ as well as the treatment of the lattice spacing scalar $a$ will not be further discussed at this point.

$(D_{nc}\psi)(x)$ is calculated on the CPU by an algorithm described in [Rot16], which is shown in Algorithm 3. As a shorthand we will refer to that algorithm as the "CPU algorithm". DDαAMG also contains a version of the algorithm that is optimised for Streaming SIMD Extensions (SSE). We will not discuss this variant in detail.

In the CPU algorithm, after the reception of $\partial_{\mu,P_i}\lambda_\mu$ and $d_{\mu,P_i}\chi_\mu$, the process has obtained the missing values from lattice sites, which are not in the local lattice. We now

**Algorithm 3:** Evaluation of the Neighbor Coupling $D_{nc}$

**Input:** Quark fields $\psi$, $\eta$
**Output:** $\eta = \eta + D_{nc}\psi$

1 **foreach** $\mu \in \mathbb{D}$ **do**
2     **foreach** $x \in P_i$ **do**
3         $\lambda_\mu(x) \leftarrow \frac{1}{2}[(I_4 - \gamma_\mu) \otimes I_3]\psi(x)$;
4     **end**
5 **end**
6 **foreach** $\mu \in \mathbb{D}$ **do**
7     start communicating $d_{\mu,P_i}\lambda_\mu$ in negative $\mu$ direction;
8 **end**
9 **foreach** $\mu \in \mathbb{D}$ **do**
10     **foreach** $x \in P_i$ **do**
11         $\chi_\mu(x + \mu) \leftarrow \frac{1}{2}[(I_4 + \gamma_\mu) \otimes U_\mu^H(x)]\psi(x)$;
12     **end**
13 **end**
14 **foreach** $\mu \in \mathbb{D}$ **do**
15     start communicating $\partial_{\mu,P_i}\chi_\mu$ in positive $\mu$ direction;
16 **end**
17 **foreach** $\mu \in \mathbb{D}$ **do**
    /* sent in Line 7                                                        */
18     wait for receiving $\partial_{\mu,P_i}\lambda_\mu$;
19 **end**
20 **foreach** $\mu \in \mathbb{D}$ **do**
21     **foreach** $x \in P_i$ **do**
22         $\eta(x) \leftarrow \eta(x) + [I_4 \otimes U_\mu(x)]\lambda_\mu(x + \mu)$;
23     **end**
24 **end**
25 **foreach** $\mu \in \mathbb{D}$ **do**
    /* sent in Line 15                                                       */
26     wait for receiving $d_{\mu,P_i}\chi_\mu$;
27 **end**
28 **foreach** $\mu \in \mathbb{D}$ **do**
29     **foreach** $x \in P_i$ **do**
30         $\eta(x) \leftarrow \eta(x) + \chi_\mu(x)$;
31     **end**
32 **end**

back-substitute $\lambda$ and $\chi$ to show how this algorithm provides us with the discretized operator $D_{nc}$.

$$\begin{aligned}
\eta(x) &\leftarrow \eta(x) + [I_4 \otimes U_\mu(x)]\lambda_\mu(x+\mu) \\
&= \eta(x) + [I_4 \otimes U_\mu(x)]\frac{1}{2}[(I_4 - \gamma_\mu) \otimes I_3]\psi(x+\mu) \\
&= \eta(x) + \frac{1}{2}([I_4(I_4 - \gamma_\mu)] \otimes [U_\mu(x)I_3]))\psi(x+\mu) \\
&= \eta(x) + \frac{1}{2}((I_4 - \gamma_\mu) \otimes U_\mu(x))\psi(x+\mu) \\
\eta(x) &\leftarrow \eta(x) + \chi_\mu(x) \\
&= \eta(x) + \frac{1}{2}[(I_4 + \gamma_\mu) \otimes U_\mu^H(x-\mu)]\psi(x-\mu)
\end{aligned}$$

When accounting for the loops over all dimensions in lines 20 and 28, as well as the additive nature of the instructions of the form $\eta(x) \leftarrow \eta(x) + *$, we obtain

$$\begin{aligned}
\eta(x) \leftarrow \ & \eta(x) + \frac{1}{2}\sum_{\mu \in \mathbb{D}}((I_4 - \gamma_\mu) \otimes U_\mu(x))\psi(x+\mu) \\
& + \frac{1}{2}\sum_{\mu \in \mathbb{D}}((I_4 + \gamma_\mu) \otimes U_\mu^H(x-\mu))\psi(x-\mu) \\
= \ & \eta(x) + (D_{nc}\psi)(x).
\end{aligned}$$

## 2.3 Storage

While in Section 1.1 we introduced a four-dimensional indexing scheme, the actual implementation must store data sequentially in memory. Enumeration of a finite set is trivial, and within the scope of our acceleration efforts the enumeration scheme is not important. We also do not use global indexing, but rather only use indices for the local lattice. As such we will consider

$$P_i \sim \{0, \ldots, |P_i|_{card} - 1\}$$
$$\overline{P_i} \sim \{0, \ldots, |P_i|_{card} - 1, |P_i|_{card}, \ldots, |\overline{P_i}|_{card} - 1\}$$

as a pragmatic formulation of the fact that each process associates each of its lattice sites with an index, which is unique in process scope. We will use these indices as if they represented an element of $\mathcal{L}$ (e.g. $\psi(i) \sim \psi(x), i \in \{0, \ldots, |P_i|_{card} - 1, |P_i|_{card}, \ldots, |\overline{P_i}|_{card} - 1\}, x \in \mathcal{L}$).

We will now introduce the storage scheme used by DDαAMG for the CPU algorithm in pseudocode, abstracting away specific types and the C code to actually allocate and construct that array. While this full introduction of the storage scheme might seem excessively detailed at this point, Section 4.5 will show that this scheme is highly detrimental to good performance on GPUs.

In order to access neighboring lattice sites in positive direction, we store the indices of neighboring sites sequentially:

$$\texttt{neighbor\_table} : \texttt{int}[4 \cdot |P_i|_{card}] = 0+T, 0+Z, 0+Y, 0+X, 1+T, \ldots$$

For the quark fields $\psi$ and $\eta$, which are also referred to as $x$ and $w$ from the FGMRES perspective, we use

$$\texttt{x} : \texttt{complex double}[12 \cdot |P_i|_{card}] = x_1(0), x_2(0) \ldots, x_{12}(0), x_1(1), \ldots$$

For the intermediate quark fields $\lambda_\mu$ and $\chi_\mu$ only six of the twelve vector entries are non-zero. Thus, only six elements per site must be stored.

$$\texttt{x}_* : \texttt{complex double}[6 \cdot |\overline{P_i}|_{card}] = x_1(0), x_2(0) \ldots, x_6(0), x_1(1), \ldots$$

For the self coupling term we gather the 42 matrix entries $c_1(x), \ldots, c_{42}(x)$ (under the misleading name clover) as

$$\texttt{clover} : \texttt{complex double}[42 \cdot |P_i|_{card}] = c_1(0), c_2(0) \ldots, c_{42}(0), c_1(1), \ldots$$

The matrices $U_\mu(x)$ have nine entries. Four of these matrices must be stored per lattice site (under the misleading name D, one for each direction):

$$\begin{aligned}
\texttt{D} : \texttt{complex double}[4 \cdot 9 \cdot |P_i|_{card}] = \\
U_{T1}(0), U_{T2}(0), \ldots, U_{T9}(0), U_{Z1}(0), \ldots, U_{X9}(0), \\
U_{T1}(1), \ldots, U_{X9}(|P_i|_{card} - 1).
\end{aligned}$$

The coefficients from the $\gamma_\mu$ matrices are preprocessor definitions of complex literals.

Let $\mathbb{E}^{n_c}$ be any $n_c$-tuple (i.e. $\mathbb{E}$ is any set of elements / is the element type). We will refer to a storage scheme $\texttt{s} : \mathbb{E}[n \cdot n_c] = s_1, \ldots, s_{n \cdot n_c}$ of an ordered set of tuples $v = \{v(1), v(2), \ldots, v(n)\}$, with $v(1), v(2), \ldots, v(n) \in \mathbb{E}^{n_c}$, as *chunkwise* iff

$$\begin{aligned}
&\forall i, j \in \{1, \ldots, n \cdot n_c\} : \\
&i < j \Rightarrow \exists k, l \in \{1, \ldots, n_c\}, p, q \in \{1, \ldots, n\} : \\
&s_i = v_k(p) \wedge s_j = v_l(q) \wedge p \leq q \wedge (p = q \Rightarrow k < l).
\end{aligned} \quad (2.8)$$

That is, the storage scheme orders and sequentializes the elements by tuples first and by tuple entry (component) second. All previously discussed arrays are using the chunkwise storage scheme. We will call the subarray of elements belonging to a single tuple a chunk.

# Chapter 3

# Performance Considerations

## 3.1 Double Precision Performance

On current NVIDIA GPUs there can be a significant difference in single and double precision computational performance. NVIDIA lists the theoretical performance for floating point operations of many datacenter products. For example, the V100 GPUs deployed in Jülich (see Section 6.1.1) have a theoretical single precision performance of 15.7 TFlop/s and 7.8 TFlop/s in double precision [NVI20]. The ratio between the two is 2:1.

For some older GPUs and also for consumer graphics cards NVIDIA does not publish this information. Various websites suggest a ratio of 64:1 for many GPU models. However, these sites do not cite any sources or methods on how this should be obtained.

Therefore, we wrote a small test application to determine this ratio on GPUs relevant to this thesis. It launches a large amount of threads with the same kernel. The kernel is completely written in inline parallel thread execution (PTX) assembly and is executing 100 floating point operations (`add.f32` or `add.f64`, respectively) sequentially (unrolled without a loop). The nvcc was instructed to omit optimizations with the `-O0 -Xptxas -O0 -Xcicc -O0` parameters. By dividing the total number of such instructions over all threads by the execution time, we get a good estimate of floating point computational performance of a GPU under unusually favorable conditions (i.e. only operating on registers). We report our results in Table 3.1.

We see a good double precision performance ratio for the V100 GPU, a current datacenter-oriented GPU model. The (older) Quadro P6000 and 3080 Ti (consumer)

| GPU | Single [GiFlop/s] | Double [GiFlop/s] | Ratio |
| --- | --- | --- | --- |
| 3080 Ti | 7357 | 246 | $\approx$30:1 |
| V100 | 6194 | 3119 | $\approx$2:1 |
| Quadro P6000 | 6345 | 187 | $\approx$34:1 |

Table 3.1: Floating Point Performance

GPUs show a large disparity in single and double precision performance. While the observed ratios of these cards rather suggest a "real" ratio of 32:1, it is not impossible that there is a different theoretical architectural ratio. Specifically, the measurement from the V100 do not even reach half of the theoretic architectural performance listed by NVIDIA in [NVI20]. There could be multiple factors that introduce this disparity: for example scheduler overheads or other bottlenecks. It is also worth mentioning that in PTX assembly there are instructions that process more than one floating point value at once e.g. the fused multiply-add (`fma`) instructions. Possibly the theoretical values consider these instructions as multiple operations. A study of Jia et al. [Jia+18] did obtain close to theoretical floating point performance. While they provide PTX assembly for most of their benchmarks, floating point performance was measured using the cuBLAS library distributed by NVIDIA. The authors state that:

»There is consensus that the high degree of optimization found in NVIDIA's libraries such as cuBlas and cuDNN is inaccessible to authors of CUDA code, even when they use inline PTX assembly. We believe that manufacturers write optimized libraries as low-level SASS code, possibly with assemblers which they do not make available to the public.«

In a pragmatic effort, we will use the ratios 32:1 and 2:1 in real world performance under unusually favorable conditions as a basis for further considerations.

## 3.2 Memory Bandwidth

A DDR4-2666 memory module runs, as the name suggests, on 2666 Mega-Transfers per second. As eight bytes are transferred on each transfer, this yields a memory bandwidth of ∼21.3 GiB/s. While most consumer CPUs have only two memory channels, server CPUs may have more than that. For example, an Intel Xeon Gold 6148 (see Section 6.1.1 for a motivation on this example) has six memory channels. So the total (theoretical) memory bandwidth in that configuration is ∼128 GiB/s.

A GPU comes with its own memory. As discussed in [Jia+18], due to the faster High Bandwidth Memory (HBM)2 used by Nvidia from generation Pascal onwards, a V100 GPU has a theoretical memory bandwidth of 900 GiB/s. In [Jia+18] researchers measured a real world bandwidth of 750 GiB/s.

Even when considering newer DDR5 memory with a doubled memory frequency (i.e. double the memory bandwidth) compared to DDR4, current GPUs outperform current CPUs in memory bandwidth. On the sparse Wilson-Dirac operator, we expect memory demand to exceed computational demand, due to the sparse structure of it, and thus memory bandwidth to be the limiting factor. In [Cla+15] it was observed that many HPC workloads are sensitive to memory bandwidth [Cla+15]. In Section 4.4 we establish that our implementation of the algorithm to apply the Wilson-Dirac operator is also memory-bound.

# Chapter 4

# GPU Port

## 4.1 Source Code Access

The DDαAMG project is developed under the GPL license version 3 and is accessible on GitHub with various forks. The original version as described in [Rot16], may be found in a git repository maintained by Rottmann[1]. The fork that works on CUDA acceleration, including the efforts from [Ram22] and this thesis, is contained in a forked repository maintained by Ramírez-Hidalgo[2].

## 4.2 General Approach

### 4.2.1 Testing

Unfortunately, the DDαAMG project does not have sufficient tests for a project of its size. Prior to the work described by this thesis there were no unit tests, only some test routines which can be enabled by undocumented compile switches. Automated integration tests did also not exist. The general approach to testing DDαAMG was and in large parts still is to use an existing linear system, create a solver configuration, run the whole DDαAMG program and try to determine whether the program behaved as expected. That is also complicated due to the fact that DDαAMG does only calculate the solution to the system involving the Wilson-Dirac operator in memory and does not provide the solution as a file or similar. There is a library version of DDαAMG which can be used to integrate it with other programs, but that was defunct (at least for the CUDA version) since the start of the work described in this thesis. Other researchers operate on forks of DDαAMG to use it in their research.

While it is clear that development is unsustainable as newly introduced bugs and regressions easily go unnoticed, a remedy (for example by introducing unit tests to the projects) is not easy. The main issue lays with the way that the program state can affect any function in the program. There is a `global_struct g` global (extern) variable that encapsulates a lot of program state. Additionally, program state is passed around in various other structs (e.g. `level_struct` or `operator_PRECISION_struct`). Therefore, functions can often not be (easily) unit-tested, as they depend on specific preconditions in so-shared program state that can not be (easily) replicated.

---

[1] https://github.com/mrottmann/DDalphaAMG
[2] https://github.com/Gustavroot/DDalphaAMG

While testing of the changes brought to DDαAMG through the described work was not broadly possible within the given timeframe, we made certain improvements. Unit testing was introduced to the project in the form of rapidcheck-/gtest tests that apply to some functions that do not depend on global state. Such tests are found under the `test/gtest` folder and are being built into the `build/gtest/dd_alpha_amg_gtest` executable with the `gtest` make recipe. An automated integration test script was also introduced that runs the DDαAMG program with various solver configurations. However, the only assertion made is that the program exits with exit code 0. A wrapper script `test_wrapper.sh` that compiles the CPU and GPU version and runs all applicable tests was added.

Otherwise, for the naive port described in Section 4.3, a temporary function that compares the results of the CPU version with the newly created GPU version of the Wilson-Dirac operator was used to validate equality. That function was removed after this first effort (i.e. for work described in sections 4.4 onwards) due to performance concerns. For subsequent work, testing was limited to checking whether the call count to the operator remained the same throughout the changes. A different call count would indicate a different result of the operator causing different convergence behavior of FGMRES.

### 4.2.2 Integrating C and CUDA C++

Portions of the code base written in CUDA C++ must be compiled with the NVIDIA-supplied nvcc compiler. The other code of the DDαAMG is written in C and typically compiled with an mpicc (gcc wrapped for MPI) compiler. This C code can not in general be compiled with a C++ compiler. A proper separation of C and CUDA C++ code is needed. This was conceptually tackled by introducing C proxy functions, wherever different or additional work needs to be performed if CUDA acceleration is enabled. Listing 4.1 shows such a function. Under the original function name `dirac_setup` the proxy function still provides the functionality for the CPU version. The original function has been renamed to `cpu_dirac_setup`. If the compilation flag `CUDA_OPT` is defined, the method to perform the additional setup required for the GPU-accelerated version of DDαAMG is called. The function `cuda_dirac_setup` must have C linkage. Currently, linking the final DDαAMG program requires the nvcc linker. This is due to an issue, where creating the `nvcc -dlink` object does not completely lift the requirement of linking relocatable device code. It is currently unclear whether this is a compiler bug or if the cause lays within DDαAMG and its build system.

Listing 4.1: Proxy Function

```
void dirac_setup(config_double hopp, config_double clover,
                 level_struct *l) {
  cpu_dirac_setup(hopp, clover, l);
#ifdef CUDA_OPT
  cuda_dirac_setup(hopp, clover, l);
#endif
}
```

### 4.2.3 Affected Source Code of DDαAMG

Up to the point where this section was written[3] the changes affected 114 source files, with 4811 insertions and 1517 deletions. A significant portion of this work came from "C++-proofing" parts of the code. This was mostly caused by the bad include structure of DDαAMG, where *all* headers were included in a single `main.h` file. These includes needed to be moved to the point where they are actually needed, such that not all code had to be compilable by a C++ compiler.

Otherwise, our changes to the source code of DDαAMG were mostly limited to introducing the proxy functions and writing of the CUDA code in the `src/gpu` directory. To investigate the changes, the `cuda_d_plus_clover_PRECISION` function is a good starting point. The work included writing operator overloads for complex CUDA numbers, extension of relevant data structures with new members (e.g. the `operator_PRECISION_struct`), some general GPU and utility functions (e.g. $3 \times 3$ matrix vector multiplication or reordering of arrays), writing of the CUDA kernels and host code to call them. For the changes in Section 4.4, some work on the communication functions was needed.

## 4.3 Naive Port

In Algorithm 3, lines 2, 10, 21 and 29 each perform an iteration over all local lattice sites. This gives rise to a basic parallelization scheme: launching a thread for each lattice site and executing the operation from the next line in parallel. We first attempted this approach in order to accelerate the application of the Wilson-Dirac operator using GPUs. This implementation is naive in the sense, that it is a very close adaptation of the CPU implementation, without any concern for the architecture of GPUs. In fact, DDαAMGs OpenMP parallelization of the discrete Wilson-Dirac operator also just splits the lattice sites between CPU threads.

As the calculations that need to be performed per lattice site were already existing as CPU functions, most of them were adaptable as kernel functions easily. A slight complication is that CUDA has its own implementation of complex numbers and their arithmetics. The C parts of DDαAMG use standardized C complex numbers. We overloaded the relevant operators for CUDA complex numbers, so the kernel functions differ less from their CPU counterparts. When we copy memory from CPU to GPU and vice versa, we rely on both implementations representing a complex number by two consecutive values in memory (i.e. `cuDoubleComplex` is defined as a struct of two double values, gcc behaves the same). We therefore do not need to convert between the two types.

---

[3]git commit 9b0e82b

The functionality of copying $\psi$ to GPU memory and $\eta$ from GPU memory is realized as a wrapper around the actual Wilson-Dirac operator function. Because data now resides in GPU memory, the vectors $\lambda_*$ and $\chi_*$ need to be copied from and to the GPU between the communication and calculation parts of the algorithm, in order to reuse existing communication functions.

## 4.4 CUDA-aware MPI Improvements

Profiling revealed that a large portion of the execution time of the naive algorithm is spent copying vectors from and to the CPU. To remedy that issue, the communication code involved needed to be rewritten. CUDA-aware MPI can handle device-allocated memory as if it was host-allocated. This can often eliminate the requirement to rewrite MPI-related code completely when porting existing C code to CUDA C++. However, the data in the vectors $\lambda_*$ and $\chi_*$ that corresponds to the inner or outer boundaries is not necessarily stored in sequence in memory. Thus, GPU buffers for communication have been added and kernels were introduced to handle reading/writing to those buffers.

The result of these improvements is that copy overhead was eliminated. Due to the structure of Algorithm 3, which hides communication behind computation, the execution time of the operator is now very close to CUDA kernel execution time. Out of the 32 ms total execution time, 31 ms are being spent in (not communication-related) kernels for the two node GMRES configuration (see also Table 6.5). This is a large improvement over the previous execution time of 128 ms.

Profiling of the kernels did reveal that all of them utilize a large portion of theoretical global memory bandwidth (around 80% of theoretical peak) and require only a small portion of the computational capabilities of the GPU (Compute Throughput is 1.9% to 2.1%). That is, application of the operator is now bottlenecked by global memory throughput.

## 4.5 Componentwise Reordering

With our implementation now being bound by global memory access, we can only improve performance if we are able to reduce the transfer volume between global memory and caches closer to our SM. We now want to discuss our memory access pattern, as NVIDIA Nsight Compute identified uncoalesced access in our kernels at this point.

Our data is stored using a chunkwise storage scheme as introduced in Section 2.3. We now walk through the kernel presented in Listing 4.2 as it was implemented naively in Section 4.3. The kernel advances the pointers `phi` and `prp_T` to the vector (the chunk) relevant to their site. As threads in a warp have consecutive indices (see calculation of `idx`), their pointers now point one chunk apart from each other (six elements

**Algorithm 4:** Evaluation of the Neighbor Coupling $D_{nc}$

**Input:** Quark fields $\psi$, $\eta$
**Output:** $\eta = \eta + D_{nc}\psi$

1 **foreach** $\mu \in \mathbb{D}$ **do**
2     **launch for** $x \in P_i$ **kernel**
3         $\lambda_\mu(x) \leftarrow \frac{1}{2}[(I_4 - \gamma_\mu) \otimes I_3]\psi(x)$;
4     **end**
5 **end**
6 **foreach** $\mu \in \mathbb{D}$ **do**
7     copy $\lambda_\mu$ from GPU to CPU;
8     start communicating $d_{\mu,P_i}\lambda_\mu$ in negative $\mu$ direction;
9 **end**
10 **foreach** $\mu \in \mathbb{D}$ **do**
11     **launch for** $x \in P_i$ **kernel**
12         $\chi_\mu(x+\mu) \leftarrow \frac{1}{2}[(I_4 + \gamma_\mu) \otimes U_\mu^H(x)]\psi(x)$;
13     **end**
14 **end**
15 **foreach** $\mu \in \mathbb{D}$ **do**
16     copy $\chi_\mu$ from GPU to CPU;
17     start communicating $\partial_{\mu,P_i}\chi_\mu$ in positive $\mu$ direction;
18 **end**
19 **foreach** $\mu \in \mathbb{D}$ **do**
    /* sent in Line 8                                                */
20     wait for receiving $\partial_{\mu,P_i}\lambda_\mu$;
21     copy $\lambda_\mu$ from CPU to GPU;
22 **end**
23 **foreach** $\mu \in \mathbb{D}$ **do**
24     **launch for** $x \in P_i$ **kernel**
25         $\eta(x) \leftarrow \eta(x) + [I_4 \otimes U_\mu(x)]\lambda_\mu(x+\mu)$;
26     **end**
27 **end**
28 **foreach** $\mu \in \mathbb{D}$ **do**
    /* sent in Line 17                                               */
29     wait for receiving $d_{\mu,P_i}\chi_\mu$;
30     copy $\chi_\mu$ from CPU to GPU;
31 **end**
32 **foreach** $\mu \in \mathbb{D}$ **do**
33     **launch for** $x \in P_i$ **kernel**
34         $\eta(x) \leftarrow \eta(x) + \chi_\mu(x)$;
35     **end**
36 **end**

Listing 4.2: Example Kernel

```
__global__ void cuda_prp_T_PRECISION(
    cu_cmplx_PRECISION * prpT,
    cu_cmplx_PRECISION const * phi,
    size_t num_sites) {
  const size_t idx = threadIdx.x + blockDim.x * blockIdx.x;
  if (idx >= num_sites){
    // there is no more site for this index
    return;
  }
  phi += 12*idx;
  prpT += 6*idx;
  prpT[0] = phi[0] -GAMMA_T_SPIN0_VAL*phi[6];
  prpT[1] = phi[1] -GAMMA_T_SPIN0_VAL*phi[7];
  prpT[2] = phi[2] -GAMMA_T_SPIN0_VAL*phi[8];
  prpT[3] = phi[3] -GAMMA_T_SPIN1_VAL*phi[9];
  prpT[4] = phi[4] -GAMMA_T_SPIN1_VAL*phi[10];
  prpT[5] = phi[5] -GAMMA_T_SPIN1_VAL*phi[11];
}
```

for `prpT`, twelve elements for `phi`). The elements are then read one after another. For a simplification Figure 4.1a sketches the access pattern as if only three threads comprise a warp.

When a value is requested (i.e. read) not only that value is loaded. Data is exchanged between the processor, caches and main (global) memory at a minimum size of a (cache) line. The line size can in general differ between the exchanging hardware elements, but that will not be relevant here. Loading a single element results in a data transfer larger (insofar the element is not a multiple of the cache line size) than the element itself. In the CUDA SIMT architecture, threads within a warp all execute the same instruction simultaneously. If all threads access consecutive elements in memory, cache lines can be used more efficiently, as the accessed elements are present consecutively in the transferred cache lines. Such an access pattern is called *coalesced memory access* in CUDA jargon [McK14, p. 159]. Because of the ambiguity whether "coalescing" means actually reducing the number of memory requests (as used in [NVI23, p. 140] and in other architectures) or using consecutive elements from the same cache line, we will not adopt that term here. We will instead refer to an access pattern where multiple threads in a warp access consecutive elements simultaneously as *thread-consecutive*. Analogously, we adopt the term *time-consecutive*, if a processing element (like a CPU core GPU thread) accesses consecutive elements sequentially.

Clearly, the `prp_T` function has a time-consecutive access pattern. Such an access pattern is unproblematic on x86-64 CPUs. As each core of an Intel Xeon Gold 6148 CPU has 64 KiB of L1 cache, the data delivered in the cache line can be worked on sequentially with minimal risk of an eviction from L1 cache. Additionally, hardware prefetchers and compiler optimization reduce latency issues for this access pattern. On

Figure 4.1: Access Patterns

(a) Chunkwise

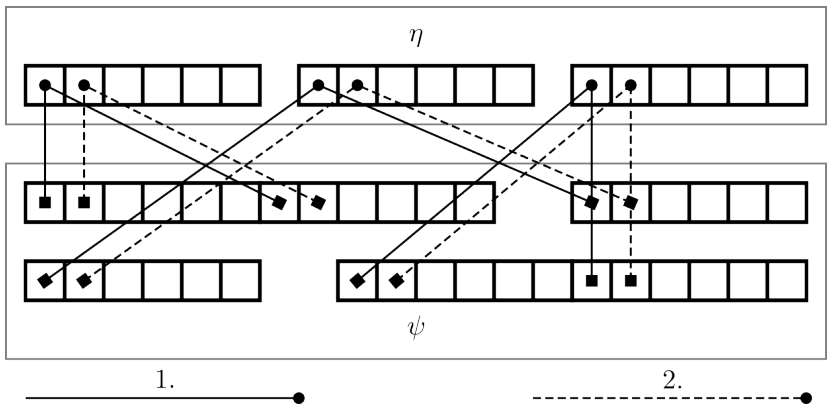

(b) Componentwise

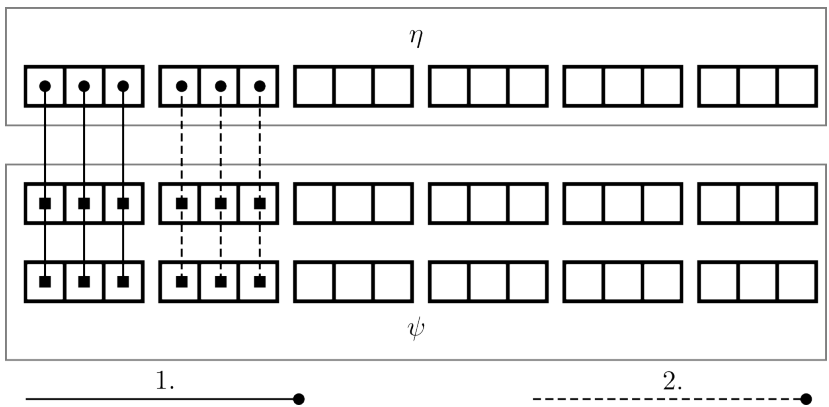

the other hand, a SM on a V100 GPU processes 32 threads together in a warp and can have up to 64 warps active at the same time (as reported by Nsight Compute). This makes a total of 2048 threads which could potentially issue memory requests. A SM has 128 KiB of L1 cache. Because only little L1 cache is available per active thread, sequential processing of the elements in a cache line is unfeasible. The time-consecutive access pattern thus needed to be changed. We considered three main options.

Firstly, there is the well-documented approach to store a copy of a memory segment, of which all values are needed, but which is not accessed thread-consecutively, in shared memory. Details on optimization with shared memory may be found in [McK14]. DDαAMG's CUDA-accelerated smoother uses shared memory in this way [Ram22, p. 92]. Early experiments with this approach proved successful and allowed to around halve the execution time of the `prp_T` kernel for large local lattices. However, usage of shared memory competes with usage of the L1 cache for actual caching. It would also introduce limits to the size of concurrently active blocks on a SM. It was not clear whether this approach would have drawbacks for some functions based on these concerns.

Secondly, it would be possible to change the kernel functions, such that their instructions access memory thread-consecutively. That is each thread only executes one of the (actual calculation) instructions from the original function, and threads perform these instructions simultaneously. We saw multiple issues with this approach. It was not clear whether warp divergence (see [McK14, p. 83]) could be avoided for all kernels. Also, from a software design standpoint it would be better to have threads handle tangible things (like lattice sites) instead of single instructions.

Thirdly, the arrays can be reordered such that thread-consecutive access is possible. That is, arrays need to be ordered by tuple entry (component) first and tuples (lattice sites) second. With the definitions from Section 2.3 we will refer to a storage scheme as componentwise iff

$$\begin{aligned}
&\forall i, j \in \{1, \ldots, n \cdot n_c\} : \\
&i < j \Rightarrow \exists k, l \in \{1, \ldots, n_c\}, p, q \in \{1, \ldots, n\} : \\
&s_i = v_k(p) \wedge s_j = v_l(q) \wedge k \leq l \wedge (k = l \Rightarrow p < q).
\end{aligned} \quad (4.1)$$

The same `prp_T` kernel would now have a thread-consecutive access pattern as seen in Figure 4.1b.

Early experiments with this approach were in general promising. The reduction in memory transfer size was large. However, also memory latency issues were observed that prevented the kernel to achieve a high memory throughput. As the componentwise reordering approach seemed especially elegant and not affected by the block restrictions of the shared memory approach, it was implemented.

We created a small wrapper class (see Listing 4.3) to simplify access to a componentwise vector. When constructed with the pointer to `ptr + idx`, it will access the components of the `idx`th vector.

Arrays are reordered from the chunkwise to the componentwise storage scheme in a way that is thread-consecutive on the destination (componentwise-ordered) array and

Listing 4.3: ComponentAccess Class

```cpp
template <typename ElementType>
class ComponentAccess {
public:
  __host__ __device__ ComponentAccess(
      ElementType* data, size_t num_sites) {
    this->data = data;
    this->num_sites = num_sites;
  }

  __device__ ElementType& operator[](size_t i) {
    return data[i * this->num_sites];
  }

private:
  ElementType* data;
  size_t num_sites;
};
```

time-consecutive on the source array. While the time-consecutive access could certainly be optimized by using shared memory, the function in Listing 4.4 is reasonably fast.

For the arrays that contain elements that need to be communicated to neighboring processes, componentwise reordering poses a problem. As the lattice sites in the inner boundaries $d_\mu P_i$ do not (in general) have consecutive indices, componentwise access would not always result in a thread-consecutive acceses pattern. While the execution time spent on copying relevant elements to a buffer would be small for large lattices, we decided to use the shared memory approach for the arrays representing $\lambda$ and $\chi$ to reduce implementation effort. In summary, the self and neighbor coupling coefficients, $\psi$ and $\eta$ are in the componentwise storage format. The arrays $\lambda$ and $\chi$ remain in chunkwise storage format and access is optimized by shared memory.

In the initial experiments on the componentwise reordering, with only some vectors reordered, latency issues were observed. In the final implementation, with all access being either thread-consecutive or buffered by shared memory, these latency issues vanished. Now, memory throughput of the kernels is measured at 80% to 90% and the transfer sizes from and to global memory are much smaller. For example, the `prpT` kernel now reads 50 MiB instead of 250 MiB from and writes 53 MiB instead of 80 MiB to global memory.

Listing 4.4: Array Reordering Function

```cpp
template <typename ElementType>
__global__ void reorderArrayByComponent(
    ElementType* dst, ElementType const* src,
    size_t chunkSize, size_t chunkCount) {
  assert(blockDim.x * gridDim.x >= chunkCount);
  const size_t idx = threadIdx.x + blockDim.x * blockIdx.x;
  if (idx >= chunkCount) {
    // there is no more chunk for this index
    return;
  }
  // set src to first element that will be read
  src += chunkSize * idx;
  auto caDst = ComponentAccess(dst + idx, chunkCount);
  for (size_t i = 0; i < chunkSize; i++) {
    caDst[i] = src[i];
  }
}
```

# Chapter 5

# Solver Configuration

In this chapter various aspects on the configuration of DDαAMG will be discussed in order to determine a good solver configuration for the measurements. As will be shown, the runtime of DDαAMG can vary widely between solver configurations even though the program solves the same linear system. We will begin with an introduction to cluster topology to ensure consistent terminology.

## 5.1 Cluster Topology

A computer cluster can be described as a collection of interconnected computers intended to collectively solve a computational task. The DDαAMG project has been designed with the execution environment and practices (e.g. job submission) of HPC clusters in mind. Nonetheless, DDαAMG does not require any specialized hardware and can run on most computer clusters insofar it offers an MPI implementation and a C compiler. The special case of a cluster made up of a single computer (i.e. running the application locally) has some relevance in the development process as changes to DDαAMG are typically tested locally first. This section will give a characterization of the structure of physical hardware in a cluster.

Within the cluster one or more computers referred to as nodes are used. The nodes have some means of communicating with each other. At Jülich Wizard for European Leadership Science (JUWELS) this is realized through an InfiniBand network [Alv22]. Within such a node one or more physical processors are located. We interchangeably also refer to the physical processor as a socket in an abstract view of the hardware. These processors share the same address space and, in principle, a program can be ignorant to its environment at that point. However, many important system components are local to each processor. This includes the main memory bus(ses), PCIe bus and other I/O busses. Access to another socket's resources is realized through a system interconnect. Latency and throughput of memory and I/O access is therefore non-uniform. This architectural pattern is thus known as Non-Uniform Memory Access (NUMA). The Linux kernel documentation on NUMA [Sar] discusses the subject in more detail. With all systems considered in Chapter 6 the terms physical processor, NUMA domain and socket refer

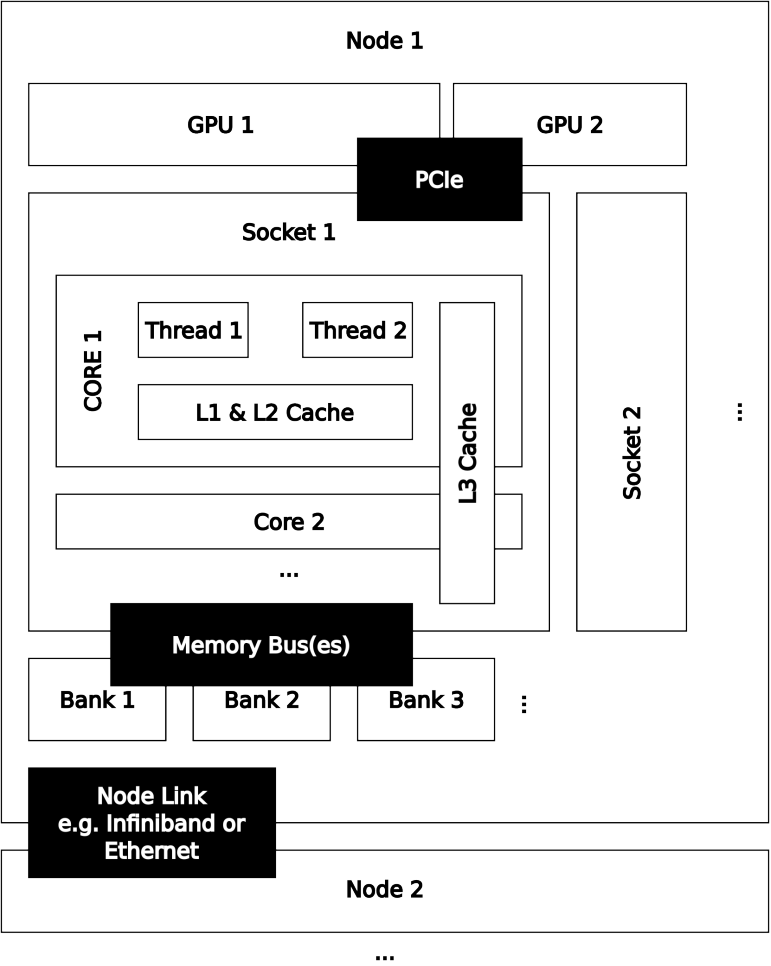

Figure 5.1: Topology of a Cluster

to the same concept.

An important issue arising from this architecture is that access to a GPU, for example to copy data from CPU to GPU memory, could result in varying throughput and latency depending on the processor accessing it. The topology between GPUs and CPUs as well as some other devices (such as InfiniBand interfaces accessible by the GPU) can be explored through the `nvidia-smi topo` command.

Each socket contains multiple cores. L3 cache is shared among those cores, but L1 and L2 caches are local to each core. These caches provide data to the cores with L1 being the fastest and L3 being the slowest cache. Note that the specifics of the cache hierarchy depend on the processor architecture. At JUWELS the cache hierarchy is as described.

Each core might support multiple (in current processors two) hardware threads. This is known as multithreading. Multithreading allows better compute throughput of the core in cases where other core resources would otherwise be underutilized.

It should be clear from this description that where processes and threads are located changes the connections between them and for which resources they compete. Also caching mechanisms do support various solver configurations differently. Due to the communication-hiding structure of the operator algorithm (see Algorithm 3), internode communication is of lesser concern than it would be for different workloads.

We also expect to be memory-bound, at least for larger lattices which do benefit less from caching. Therefore, we limited the task of finding a good solver configuration mostly to balancing overheads for processes and threads, memory utilization and caching. Avoiding higher latency and lower throughput, due to data needing to traverse the system interconnect, shall also be considered.

## 5.2 Mapping & Binding

The MPI 4.0 standard [Mes21] defines many aspects of how MPI processes can communicate with each other and to some extent how they can discover their environment and the process topology. MPI even allows to define a "virtual topology" that arranges the processes in a format consistent with the problem or algorithm.

The standard is also very clear about it not defining mechanisms to assign processes to physical hardware. For example, it is stated that "though physical mapping is not discussed, the existence of the virtual topology information may be used as advice by the runtime system" [Mes21, p. 389], "MPI is primarily concerned with communication rather than process or resource management" [Mes21, p. 487] and that "[source code portability] explicitly does not say anything about how an MPI program is started or launched from the command line, nor what the user must do to set up the environment in which an MPI program will run". [Mes21, p. 487].

The task of spawning the processes and establishing communication links between them is implementation-dependent. The standard often refers to the system that handles these tasks as the execution environment or runtime. In existing MPI implementations the task of spawning processes in multiple locations and the communication library are typically split. For example, the MPI implementation OpenMPI does provide the communication library and by default comes with `orterun` as a tool to spawn the processes.

OpenMPI can also be combined with the Slurm resource manager to spawn processes. We will now adopt the terminology of calling the library (or libraries) linked to handle calls to MPI standard-defined functions as the "MPI communication library" and the tool that spawns the processes in multiple locations as the "process spawner".

A lot of functionality of the process spawners is comparable. We will however, focus on Slurm, because it offers more sophisticated options with respect to mapping and binding to GPU and CPU resources and because it is the process spawner at JUWELS.

Descriptions of some experiments to determine a good solver configuration may be found in Annex A. The MPI communication library for all experiments at JUWELS will be ParaStation MPI 5.7.0. The experiments are performed in the (lesser utilized) `develgpus` partition which offers two nodes of the same specification as described in 6.1.1. We will perform these experiments with the CPU version of DDαAMG without SSE acceleration. The description of the experiments will follow a format where a hypothesis is given, then two or more solver configurations to test the hypothesis, then an expected result if the hypothesis would be true, and finally the measured result and a discussion of the result.

## 5.3 Solver Configurations for Measurements

For reproduction purposes the DDαAMG configurations and batch job scripts that have been used for the measurements in Chapter 6 are available on GitHub [1] under the EUPL-1.2 open source license. In the same repository, exact solver configurations for the experiments in Annex A may be found. The large $64 \times 32^3$ gauge configuration can be provided on request, as it does not easily fit in a git repository.

The solver configurations for our measurements differ in two options. The first option is the size of the global and local lattice and thus the count of spawned processes. The second option is whether AMG preconditioning is applied. We ask for a reduction of the initial residual by a factor of $10^{-10}$ ($tol$ value) throughout all solver configurations.

The sizes of the local lattices are given by Table 5.1. With the tiny configurations, we can measure the performance of the application for small local lattices without using a large cluster (eight processes on two nodes). The two node configurations use significantly larger global and local lattices, but also spawn eight processes on two nodes. Through these configurations we can measure performance on relatively large local lattices. The large grid configuration solves the same linear system, but uses 16 instead of two nodes. While we would have preferred a much larger cluster (possibly introducing larger communication overheads to study), our project was restricted to the JUWELS GPU partition, which only offers 46 nodes (as reported by `sinfo` contrary to the documentation [JUW22]). Out of those 46 nodes, 28 were drained (not accepting jobs), when we performed our measurements.

The size of the global and local lattices at depth 1 for the AMG configurations are given by Table 5.2. Since tweaking of AMG parameters is not in scope, we arbitrarily chose 24 test vectors and 4 setup iterations for all AMG solver configurations.

---

[1] https://github.com/tmattha/alpha-repro

| Configuration | Global | Local | $m_0$ | $c_s w$ |
|---|---|---|---|---|
| tiny | $8^4$ | $4^3 \times 8$ | -0.1 | 1.0 |
| two node | $64 \times 32^3$ | $32 \times 16 \times 16 \times 32$ | -0.332 | 1.0 |
| large grid | $64 \times 32^3$ | $16 \times 16 \times 8 \times 16$ | -0.332 | 1.0 |

Table 5.1: Configuration Lattice Sizes

| Configuration | Global | Local | test vectors | setup iterations |
|---|---|---|---|---|
| tiny | $2^4$ | $1^3 \times 2$ | 24 | 4 |
| two node | $8 \times 4^3$ | $4 \times 2 \times 2 \times 4$ | 24 | 4 |
| large grid | $16 \times 8^3$ | $4 \times 4 \times 2 \times 4$ | 24 | 4 |

Table 5.2: Depth 1 AMG Lattice Sizes

Because of our observations from Sections A.4 and A.1, we try to use as few processes as possible while spawning at least one process on each socket. A solver configuration of one process per socket would be possible for the CPU and SSE jobs. To fully utilize the available GPU resources at JUWELS, we require two processes per socket (one per GPU). In order to avoid different local lattice sizes, we compromise in favor of two processes per socket for the CPU and SSE jobs as well. The binding of GPU resources to processes in the same NUMA domain is automatically handled by Slurm. Thus, each process obtains one GPU for exclusive use.

As suggested by the observations in Section A.2, we intentionally do not use hyperthreading. Ten threads are being bound to each process. As the number of bound threads is exactly half the core count per socket, the two processes on each socket split all cores of the socket between each other equally. This is ensured through the `--distribution=block:cyclic:fcyclic` flag. With this option the processes also do not distribute their threads over multiple sockets, which would result in worse performance as determined in Section A.3.

# Chapter 6

# Performance Measurements

In this chapter we will present our performance measurements in order to evaluate our GPU acceleration efforts.

## 6.1 System Specifications

### 6.1.1 JUWELS

Performance tests were performed on the JUWELS supercomputer at Forschungszentrum Jülich. Specifically, nodes from the GPU partition of the JUWELS cluster were used which offer two CPUs and four GPUs per node. The details of the node specifications are described in [Alv22], and Table 6.1 gives an overview.

The V100 GPUs in the JUWELS GPU partition are datacenter-oriented components without any display connectors. The main difference to consumer GPUs for the port of the Wilson-Dirac operator is that they offer good performance for double precision operations (see Section 3.1). They feature NVLink connections between GPUs.

### 6.1.2 Internal Testing

Internal tests at the Department of Mathematics & Informatics were performed on a machine that is called aicomp04. Aicomp04 is physically located at the University of Wup-pertal. Its specifications are listed in Table 6.2. While the main focus of the DDαAMG project is to obtain good performance on HPC clusters, aicomp04 provides us with another environment to study the impact of our acceleration efforts. The Quadro P6000 GPUs only offer around half the memory bandwidth of the V100s in Jülich [NVI16] and worse double performance as established in Section 3.1. We did not repeat our final measurements on this machine, but our results from Section 3.1 were obtained here.

| CPUs | $2 \times$ Intel Xeon Gold 6148 |
|---|---|
| System Memory | 192 GiB |
| GPUs | $4 \times$ NVIDIA Tesla V100-SXM2-16GB |
| GPU global memory | 16 GiB memory |
| GPU memory bandwidth | 900 GiB/s [NVI20] |
| CUDA version | 12.0 |
| GPU Driver version | 525.85 |

Table 6.1: GPU Node Specification at JUWELS

| | |
|---|---|
| CPUs | 2 × Intel(R) Xeon(R) Platinum 8180 |
| System Memory | 1.5 TiB |
| GPUs | 3 × Nvidia Quadro P6000 |
| GPU global memory | 24 GiB memory |
| GPU memory bandwidth | 432 GiB/s [NVI16] |
| CUDA version | 10.2 |
| GPU Driver version | 440.44 |

Table 6.2: aicomp04 Specification

## 6.2 Measurement Method

There is already a profiling implementation present in DDαAMG, which was reused to obtain the performance measurements in this chapter. It will not be discussed in great detail, but conceptually it measures differences in `MPI_Wtime()` (the MPI function to obtain a wall clock time) and records call counts. This yields a pair of total wall clock time spent on a section and a call count. `MPI_Wtick()` reports a clock resolution of 1 ns. However, measurements differ between runs by significantly more than a couple of nanoseconds. There are multiple potential sources for these measurement variations. Hardware effects could be significant, for example thermal throttling on long-running tasks. Also, aicomp04 is a multi-user machine and exclusive use is not guaranteed. At JUWELS exclusive use of nodes is guaranteed through the batch system, so resource competition is rather minimal (except with system background tasks). Variations are, as expected, thus smaller at JUWELS.

The reader is advised to act with sufficient skepticism regarding these measurements. Somewhat arbitrarily, a precision of 2 digits before cutoff (e.g. 7259.1 → 7300 or 1.346 → 1.3) was chosen for the measurements to not give a misleading impression of the accuracy of a measurement. Average time to complete a section is calculated by dividing total time by call count. The best run was used, where multiple runs have been performed. This reflects our belief that worse results do not indicate worse performance of the implementation but are rather caused by external factors.

## 6.3 Baseline

As a first step, baseline results were collected for the performance of the operator for DDαAMG on CPUs at the JUWELS cluster. We present the results in Tables 6.3 and

Table 6.3: Baseline Operator Performance Measurements

| Configuration | Count | Self Coupling | | Neighbor Coupling | |
|---|---|---|---|---|---|
| | | Total [ms] | Avg. [ms] | Total [ms] | Avg. [ms] |
| tiny CPU GMRES | 96 | 0.82 | 0.0085* | 10 | 0.10 |
| tiny SSE GMRES | 96 | 0.80 | 0.0083* | 8.0 | 0.083 |
| two node CPU GMRES | 124 | 800 | 6.5 | 3500 | 28 |
| two node SSE GMRES | 124 | 810 | 6.6 | 3400 | 28 |
| large grid CPU GMRES | 124 | 100 | 0.83 | 480 | 3.9 |
| large grid SSE GMRES | 124 | 88 | 0.71 | 450 | 3.7 |
| tiny CPU AMG | 7 | 0.11 | 0.01 | 1.2 | 0.17 |
| tiny SSE AMG | 7 | 0.07 | 0.01 | 1.2 | 0.17 |
| two node CPU AMG | 8 | 53 | 6.6 | 240 | 30 |
| two node SSE AMG | 8 | 54 | 6.7 | 260 | 32 |
| large grid CPU AMG | 9 | 8.2 | 0.91 | 47 | 5.3 |
| large grid SSE AMG | 9 | 8.7 | 0.97 | 42 | 4.6 |

* The second (non-zero) digit of the measurement is below clock resolution.

Table 6.4: Baseline Solve Time Measurements

| Configuration | setup | solve [ms] |
|---|---|---|
| tiny CPU GMRES | - | 52 |
| tiny SSE GMRES | - | 37 |
| two node CPU GMRES | - | 19000 |
| two node SSE GMRES | - | 19000 |
| large grid CPU GMRES | - | 1800 |
| large grid SSE GMRES | - | 2500 |
| tiny CPU AMG | 790 ms | 44 |
| tiny SSE AMG | 410 ms | 26 |
| two node CPU AMG | 140 s | 5000 |
| two node SSE AMG | 59 s | 4300 |
| large grid CPU AMG | 19 s | 710 |
| large grid SSE AMG | 5600 ms | 470 |

6.4. While such results were previously obtained and reported in publications [Fro+14] and [Ram22], new measurements were taken because of three reasons. Firstly, to reproduce these results in general and also to determine if they are reproducible after the various changes that have been made to the DDαAMG project. Secondly, the authors of [Fro+14] and [Ram22] did not publish all details about their configurations and a consistent configuration is needed to compare to these baseline results. Thirdly, the results in [Fro+14] were obtained from the older Juropa supercomputer at Jülich, which was decommissioned in 2015.

The improvements through SSE instructions reported in [Rot16] were in general reproducible on AMG configurations. For a two level method a reduction of 20% in solve time was reported in [Rot16, p. 103]. Due to the long reported solve times with large number of processes, exclusive use of processes for parallelization in [Rot16] seems likely. The configurations we carefully chose throughout Chapter 5 make heavy use of OpenMP threads instead. The closest match for the configuration used for evaluation of SSE in [Rot16, p. 103] for the two grid scenario is our *two node CPU/SSE AMG* configuration. While the system to solve and the AMG configuration have large differences, we use 80 processing elements from two nodes to solve a $64 \times 32^3$ system and the mentioned configuration uses 128 processes to solve a $64^4$ system. Work per core is thus in a similar range.

The results from Table 6.4 show a reduction of 15% in solve time for the SSE-optimized two node AMG configuration. All differences between configurations considered, this is quite consistent with the measurements from [Rot16, p. 103]. We did however, observe larger differences in setup time throughout most configurations. The SSE version reduces setup time by 59% for the two node AMG configuration compared to the original CPU version.

Table 6.3 shows that SSE has little effect on the execution time spent on the Wilson-Dirac operator, with some measurements even showing increased execution times.

We believe the finding that SSE has little effect on the execution time of the discrete Wilson-Dirac operator, but results in large reductions of total execution time for AMG configurations to indicate the following. The implementation of the CPU algorithm to apply the discrete Wilson-Dirac operator is memory-bound. Use of more efficient SSE instructions does not result in better performance, as data is not available fast enough. On the other hand, there are parts of DDαAMG that are not memory-bound, most notably the SAP smoother. These benefit from the more efficient SSE instructions.

The AMG preconditioning did not pay off for any of our configurations. However, the system for the *two node* and *large grid* configurations is not especially large or ill-conditioned. We also did not spend any effort adjusting multigrid parameters. Tuning, especially of the coarse grid size and number of test vectors, could yield very different results. Because a thorough study of multigrid parameters does not contribute anything to the topic of this thesis, this issue can't be discussed in large detail. However, the choice of a smaller $4 \times 2^3$ depth 1 grid and a reduction of the number of test vectors to 6, did produce a much faster AMG configuration for the two node system in local experiments.

The time spent on the application of the Wilson-Dirac operator represents a larger share of total solve time for a pure GMRES method (∼23% for the two node GMRES configuration) than for an AMG method (∼6% for the two node AMG configuration).

Table 6.5: GPU Operator Performance Measurements

| Configuration | Count | Self Coupling | | Neighbor Coupling | |
|---|---|---|---|---|---|
| | | Total [ms] | Avg. [ms] | Total [ms] | Avg. [ms] |
| tiny naive GMRES | 96 | 8.9 | 0.94 | 81 | 0.84 |
| tiny cu.a. GMRES | 96 | 9.1 | 0.095 | 46 | 0.47 |
| tiny c.w. GMRES | 96 | 8.2 | 0.086 | 47 | 0.49 |
| two node naive GMRES | 124 | 1000 | 8.2 | 14 000 | 120 |
| two node cu.a. GMRES | 124 | 1000 | 8.2 | 2900 | 24 |
| two node c.w. GMRES | 124 | 200 | 1.6 | 850 | 6.9 |
| large grid c.w. GMRES | 124 | 28 | 0.23 | 170 | 1.4 |
| tiny c.w. AMG | 7 | 4.8 | 0.69 | 9.8 | 1.4 |
| two node c.w. AMG | 8 | 17 | 2.1 | 70 | 8.7 |
| large grid c.w. AMG | 8 | 6.1 | 0.76 | 24 | 3.0 |

Table 6.6: GPU Solve Time Measurements

| Configuration | setup | solve [ms] |
|---|---|---|
| tiny naive GMRES | - | 110 |
| tiny cu.a. GMRES | - | 80 |
| tiny c.w. GMRES | - | 82 |
| two node naive GMRES | - | 32000 |
| two node cu.a. GMRES | - | 21000 |
| two node c.w. GMRES | - | 18000 |
| large grid c.w. GMRES | - | 1600 |
| tiny c.w. AMG | 550 ms | 50 |
| two node c.w. AMG | 16 s | 1100 |
| large grid c.w. AMG | 3047 ms | 47 |

## 6.4 GPU results

We present the results of our GPU measurements in Tables 6.5 and 6.6. At first, we would like to discuss the results from the GMRES configurations. The AMG results are heavily influenced by the previous work in [Ram22].

Clearly the naive GPU port did perform worse than the CPU and SSE versions of DDαAMG. The large overhead of the memory copies over PCIe lead to more execution time being spent on the operator.

The use of CUDA-aware (shorthand cu.a.) MPI did yield much better results. In the two node GMRES configuration, this version performed close to the CPU and SSE version (spent 1000 ms vs. 800 ms on self coupling, 2900 ms vs. 3500 ms on neighbor coupling).

The componentwise reordering version from Section 4.5, is another large improvement. It yields a speedup for the self coupling part of 5 over the CUDA-aware MPI version. The speedup for the neighbor coupling is around 2.9. With the additional work to reorder the arrays `phi` and `eta`, the total solve time is only slightly lower with 18 s instead of 19 s. It shall be noted, that application of the discrete Wilson-Dirac operator only takes a small share of total solve time of around 1 s compared to the CPU versions 4.3 s in the two node GMRES configuration. So ultimately, if, as suggested in [Ram22], the whole finest level is calculated on GPUs and overheads are thus further reduced, our acceleration efforts could contribute to an overall faster solver.

For small local lattices (i.e. the tiny configurations), we did not achieve improvements over the CPU or SSE version. CUDA-related overheads leave us with a longer solve time.

We now want to get to the discussion of the AMG results. For small local lattices the componentwise reordered GPU version is between the CPU and SSE version in total execution time. For large local lattices, we see large reduction in both setup time and solve time. Notably, the setup of the two node c.w. AMG configuration is ∼8.8 times faster than the two node CPU AMG configuration (16 s vs. 140 s). The setup only involves the GPU acceleration of the SAP smoother. Curiously, [Ram22] did not find a speedup in total execution time, but a large speedup of more than 30 for the actual SAP computations. The choice of a large number of nodes (with a total of 256 GPUs) and relatively small local lattices within that thesis seems to have hidden the considerable acceleration that was achieved. Overall, the current AMG GPU version of DDαAMG, with the improvements from [Ram22] as well as this thesis, outperforms the AMG CPU and SSE versions for medium and large lattices.

# Chapter 7

# Conclusion

We determined that our implementation of the discrete Wilson-Dirac operator is memory bound. Our GPU implementation of the operator achieves a speedup of ~4 against the original CPU version for the largest local lattices we measured (the two node configuration). This speedup mostly comes from the faster speed of the HBM2 modules on the V100 GPUs at the JUWELS supercomputer. We only use a small share of the computational resources of the GPU. Due to the low share of work spent on the discrete Wilson-Dirac operator and the additionally introduced overheads, the total solve time did not decrease to a significant extent. On small local lattice configurations, our GPU implementation performs similar or even worse than the original CPU version. The GPU port required large changes to the source code of DDαAMG. For a fast implementation, we needed to rewrite communication functions to make use of CUDA-aware MPI. We also reordered our arrays for more efficient use of memory bandwidth.

This is a much different result than what was achieved through the GPU acceleration of the SAP in [Ram22]. Even though, in the original doctoral thesis, an increase in total execution time was reported, we determined that for larger local lattices, these improvements yield large reductions in total execution time. The CPU implementation of SAP was not memory-bound, but rather compute bound. Thus, the GPU implementation was able to take advantage of the GPU compute resources.

We thus conclude, that GPU acceleration is a useful tool to accelerate compute-bound portions of the DDαAMG solver for large local lattices. For memory-bound portions, we did find that our improvements did not outweigh the additional overheads we introduced.

## 7.1 Future Work

Before porting more of DDαAMG to the GPU, the current AMG implementation should be extensively profiled. Collecting performance metrics may help to identify portions, which are compute-bound and deal with large problem sizes. These portions would likely be better candidates for GPU acceleration than the discretized Wilson-Dirac operator covered in this thesis.

Faster memory will likely not stay the privilege of GPU architectures. Intel already offers x86-64 CPUs with HBM2 in its *Max* series processors (e.g. the Intel Xeon Max 9462 CPU). For the memory-bound portions of DDαAMG, such CPUs could in theory offer the same improvements, which we obtained through the work described in this thesis, without introducing the same overheads. Monitoring these architectural advancements and their adoption in HPC clusters is advised.

For the DDαAMG project, we must also decide, whether GPU acceleration shall be a research topic going forward. The restrictiveness of the local lattice sizes that benefit from GPU acceleration, definitely poses some problems. Alternative optimization routes would rather be algorithmically motivated. For example, a study of the error propagator of the AMG preconditioner or heuristics for a choice of good multigrid parameters could be interesting starting points. Also, parallel solving for multiple right-hand sides could give raise to more efficient algorithms.

# Appendix A

# Experimental Solver Configuration Results

## A.1 Processes on new Sockets

**Hypothesis**

A larger number of processes improves performance if they are launched on a new socket.

**Solver Configurations**

Configuration A uses `--ntasks-per-node=1` and a local lattice of size $32^4$.
Configuration B uses `--ntasks-per-node=2`, `--ntasks-per-socket=1` and a local lattice of size $16 \times 32^4$.

**Expected Behavior**

Configuration B outperforms A.

**Result & Discussion**

| Configuration | neighbor coupling [s] | FGMRES [s] |
|---|---|---|
| A | 49 | 280 |
| B | 25 | 141 |

Configuration B did outperform A. As execution time decreased by approximately half, it also seems likely that previously completely unused resources have been utilized. Possible increased communication overheads do not seem to affect the result significantly.

## A.2 Core Performance

**Hypothesis**

More assigned hardware threads with the corresponding number of threads improve performance up to an upper limit given by the number of hardware threads of the system.

**Solver Configurations**

All configurations use `--ntasks-per-node=2`, `--ntasks-per-socket=1` and a local lattice of size $16 \times 32^3$.
Configuration A uses `--cpus-per-task=10`.
Configuration B uses `--cpus-per-task=20` (one thread per core).
Configuration C uses `--cpus-per-task=40` (one thread per hardware thread).
Configuration D uses `--cpus-per-task=80` and `--oversubscribe`.

The number of OpenMP threads is set to the same value.

**Expected Behavior**

Configuration C performs better than configurations A and B as well as better or equal to configuration D.

**Result & Discussion**

| Configuration | neighbor coupling [s] | FGMRES [s] |
| --- | --- | --- |
| A | 5.1 | 27 |
| B | 3.4 | 18 |
| C | 3.4 | 19 |
| D | 3.6 | 18 |

Configuration B, C and D had very similar execution times. Configuration A did perform worse. There does not seem to be any advantage from hyperthreading. While the worse performance of configuration A could indicate that we are compute-bound, the results for configurations B to D could also indicate better usage of L1 cache which is local to each core, and DDαAMG is still bound by access to main memory. Under the second assumption, the effect of the `--hint=memory_bound` Slurm option, which is that only one core in each socket is used, yields worse results for memory-bound applications than a configuration using all cores (e.g. `--hint=compute_bound`). The results are not consistent with the hypothesis as an upper limit is already reached under configuration B (one thread per core).

## A.3 Processes over multiple Sockets

**Hypothesis**

Distributing the threads of multiple processes over multiple sockets achieves worse performance than if the processes each used one socket exclusively.

**Solver Configurations**

Both configurations use `--ntasks-per-node=2`, a local lattice of size $16 \times 32^3$ and `--cpus-per-task=20`. The number of OpenMP threads is set to the same value.

Configuration A uses `--distribution=block:fcyclic:fcyclic` (spreads threads over both sockets in a node).

Configuration B uses `--distribution=block:cyclic:fcyclic` (keeps threads of each process on a single socket).

**Expected Behavior**

Configuration B outperforms configuration A.

**Result & Discussion**

| Configuration | neighbor coupling [s] | FGMRES [s] |
|---|---|---|
| A | 4.3 | 23 |
| B | 3.4 | 18 |

The results are consistent with the hypothesis.

## A.4 Process Costs

**Hypothesis**

It does not affect the execution time whether multiple processes or the same number of threads are used.

**Solver Configurations**

Configuration A uses `--ntasks-per-node=32`, a local lattice of size $4 \times 8 \times 32^2$ `--cpus-per-task=1`.

Configuration B uses `--ntasks-per-node=2`, a local lattice of size $16 \times 32^3$ and `--cpus-per-task=16`.

**Expected Behavior**

Configurations A and B perform comparably.

**Result & Discussion**

| Configuration | neighbor coupling [s] | self coupling [s] | FGMRES [s] |
|---|---|---|---|
| A | 7.1 | 1.6 | 37 |
| B | 3.8 | 0.98 | 20 |

Configuration A performed significantly worse than configuration B. This is inconsistent with the hypothesis. It is noteworthy that execution time of the self coupling section (i.e. the Clover term) did also increase. The clover term does not depend on values outside the local lattice. This suggests that the increased execution time does not stem from the different splitting and relatively larger halos that result from smaller local lattices. Context switching is also not expected to a large extent as processes are each pinned to a single core and nodes at JUWELS do not run a lot of background processes. Some other indicators are given by the `perf` profiling tool. `perf stat -e cache-misses,cache-references,instructions,cycles` reveals cache miss rates of 96 to 97% for processes from configuration A, but only 92% for configuration B. Configuration A achieves instructions per cycle down to 0.56, while configuration B does achieve relatively consistent results for all processes of 0.80 instructions per cycle. However, the better performance of configuration B has not been fully explained at this point.